\documentclass[aps,prd,preprint,11pt, onecolumn, tightenlines, notitlepage, superscriptaddress, nofootinbib, preprintnumbers, floatfix]{revtex4}
\pdfoutput=1

\usepackage[x11names]{xcolor}
\usepackage[colorlinks]{hyperref}

\usepackage{amstext}
\usepackage{amssymb}
\usepackage{amsmath}
\usepackage{graphicx}
\graphicspath{{plots/}}
\usepackage{hyperref}
\usepackage{url}
\usepackage{color}
\usepackage[normalem]{ulem}
\usepackage[utf8]{inputenc}
\usepackage{textcomp}
\usepackage{todonotes}
\usepackage{cancel}

\newcommand {\ignore}[1]{}
\usepackage{epsfig,amsfonts,mathrsfs,slashed,multirow}
\usepackage{latexsym,enumerate,cancel,gensymb}
\definecolor{darkred}{rgb}{0.6,0,0} %%% jv
\definecolor{mightnightblue}{RGB}{25,25,112}
\definecolor{brown}{rgb}{0.59, 0.29, 0.0}

\def\Valencia{Departament de F\'isica  Te\'orica,  Universitat  de  Val\`{e}ncia, and Instituto de F\'{i}sica Corpuscular, CSIC-Universitat de Val\`{e}ncia, 46980 Paterna, Spain}

\AtBeginDocument{\hypersetup{citecolor=purple,linkcolor=purple,urlcolor=purple}}

\begin{document}

\title{\Large Non-standard interactions from the future neutrino solar sector }

\author{P. Mart\'{i}nez-Mirav\'e}\email{pamarmi@ific.uv.es}
\affiliation{\Valencia}
\author{S. Molina Sedgwick}\email{susanams@ific.uv.es}
\affiliation{\Valencia}
\author{M. T{\'o}rtola}\email{mariam@ific.uv.es}
\affiliation{\Valencia}

\begin{abstract}
\vskip .5cm
The next-generation neutrino experiment JUNO will determine the solar oscillation parameters - $\sin^2 \theta_{12}$ and $\Delta m^2_{21}$ - with great accuracy, in addition to measuring $\sin^2\theta_{13}$, $\Delta m^2_{31}$, and the mass ordering. In parallel, the continued study of solar neutrinos at Hyper-Kamiokande will provide complementary measurements in the solar sector. In this paper, we address the expected sensitivity to non-universal and flavour-changing non-standard interactions (NSI) with $d$-type quarks from the combination of these two future neutrino experiments. We also show the robustness of their measurements of the solar parameters $\sin^2 \theta_{12}$ and $\Delta m^2_{21}$ in the presence of NSI. We study the impact of the exact experimental configuration of the Hyper-Kamiokande detector, and conclude it is of little relevance in this scenario. Finally, we find that the LMA-D solution is expected to be present if no additional input from non-oscillation experiments is considered.

\end{abstract}

\maketitle
\newpage
\tableofcontents
\section{Introduction}
The three-neutrino oscillation picture is well-established from long-running studies of solar, reactor, atmospheric and accelerator neutrinos. This quantum phenomenon can be parametrised in terms of two mass splittings ($\Delta m^2_{21}$ and $\Delta m^2 _{31}$), three mixing angles ($\theta_{12}$, $\theta_{13}$ and $\theta_{23}$) and a phase accounting for CP non-conservation ($\delta_{CP}$). Measurements of these parameters are now entering a precision era, with three of them ($\Delta m^2_{21}$, $\theta_{12}$ and $\theta_{13}$) already unambiguously determined to the percent level~\cite{deSalas:2020pgw, Esteban:2020cvm, Capozzi:2018ubv}. Our knowledge is based on the complementarity between experiments, which has also helped to shed light on two remaining open issues - the octant of $\theta_{23}$ and the mass ordering (in other words, the sign of $\Delta m^2_{31}$). Nonetheless, there is a tension between recent results from T2K and NO$\nu$A on the value of $\delta_{CP}$ for the preferred mass ordering~\cite{Abe:2021gky,Kolupaeva:2020pug}. These and other open questions will be addressed in both current and next-generation neutrino experiments~\cite{tortola_mariam_2018_1286935,josh_klein_2020_4140246, masaki_ishitsuka_2020_4142049, michael_mooney_2020_4134669}.

The discovery of flavour oscillations provided the first clear evidence that neutrinos are massive particles and as such, that the Standard Model (SM) as it stands is not enough to explain the nature of particle physics in its entirety, as it does not provide a clear mechanism for the origin of their mass. A viable dark matter candidate has also yet to be found, though its nature and potential connection to neutrinos is a matter of extensive study.

Many extensions of the SM which attempt to address these unsolved puzzles share a common feature: the emergence of effective non-standard interactions (NSI) between neutrinos and SM fields~\cite{Valle:1990pka,Ohlsson:2012kf,Miranda:2015dra,Farzan:2017xzy}. If such interactions were strong enough with respect to those in the SM, they would be detectable in neutrino oscillation and neutrino scattering experiments. In general, non-standard interactions can be divided into Charged-Current NSI (CC-NSI) and Neutral Current NSI (NC-NSI). While CC-NSI are only relevant in neutrino production and detection processes, NC-NSI can affect neutrino detection as well as their propagation in matter. In addition, non-standard interactions provide additional sources of CP violation, a topic that has been the subject of renewed attention following the recent tension in measurements carried out by NO$\nu$A and T2K~\cite{Denton:2020uda,Chatterjee:2020kkm}.

So far there has been no evidence for the existence of NSI and, therefore, we have only upper bounds on the strength of these new interactions coming from different types of experiments (for a comprehensive review on the status of NSI we refer the reader to Ref.~\cite{Farzan:2017xzy}).
Further improvements in the constraints on NSI are expected from the next generation of oscillation experiments, since they will be sensitive to subleading effects. In the short-term, the medium-baseline reactor experiment JUNO~\cite{Djurcic:2015vqa} will measure the oscillation parameters of the solar sector $\theta_{12}$ and $\Delta m^2_{21}$, together with $\Delta m^2_{31}$, with unprecedented accuracy~\cite{An:2015jdp, Abusleme:2021zrw}. 
 In parallel, the Hyper-Kamiokande detector~\cite{Hyper-Kamiokande:2018ofw} will study the high-energy spectrum of solar neutrinos with a vastly increased statistical power with respect to its predecessor, Super-Kamiokande \cite{Abe:2018uyc}. 
The complementarity of solar and long-baseline reactor experiments has been shown to successfully curtail the existence of non-standard interactions in the past~\cite{Miranda:2004nb, Escrihuela:2009up, Friedland:2004pp, Palazzo:2011vg, Guzzo:2004ue} and it will continue to be explored in the future.

In this paper, we address the expected sensitivity of JUNO and Hyper-Kamiokande to NC-NSI with $d$-type quarks\footnote{The result for $u$-type quarks can be adapted by correcting the different fraction within the Sun. For electrons, however, the analysis would  be more complex, since NSI would also affect the neutrino-electron scattering detection process in Hyper-Kamiokande and, therefore, axial NSI couplings would need to be considered as well. A complete analysis - including flavour-changing as well as non-universal vectorial and axial NSI couplings with electrons - would reach a considerable level of complexity and the large number of degrees of freedom would result in poor sensitivity to individual NSI couplings. In addition, the study of NSI with electrons is less phenomenologically interesting than that of NSI with quarks, since the degenerate LMA-D solution is completely excluded in the former due to the larger values required for non-universal NSI couplings.} and test the robustness of their measurements of solar oscillation parameters in the face of NSI. In section \ref{sec:teo}, we introduce the effective formalism used to parametrise NSI and how it translates to an effective two-neutrino approach. In sections \ref{sec:hk} and \ref{sec:juno}, the methods used in the simulation of both Hyper-Kamiokande and JUNO are explained in detail, and their individual sensitivities in the absence of NSI are examined. In section \ref{sec:results}, the sensitivity of each experiment to NSI is presented, as well as the expected results from the combination of both experiments. We also comment in this section on the status of the so-called LMA-D solution~\cite{Miranda:2004nb}, based on neutrino oscillation experiments alone. Finally, our main conclusions are summarised in section \ref{sec:conclusion}.

%%%%%%%%%%%%%%%%%%%%%%%%%%%%%%%%%%%%%%%%%%%%%%
%%%%%%%%%%%%%%%%%%%%%%%%%%%%%%%%%%%%%%%%%%%%%%
\section{NSI and their impact on neutrino oscillations}
\label{sec:teo}
%%%%%%%%%%%%%%%%%%%%%%%%%%%%%%%%%%%%%%%%%%%%%%
%%%%%%%%%%%%%%%%%%%%%%%%%%%%%%%%%%%%%%%%%%%%%%
\subsection{General formalism}
%%%%%%%%%%%%%%%%%%%%%%%%%%%%%%%%%%%%%%%%%%%%%%

Non-standard interactions can be studied within the frame of effective field theories through their parametrisation in terms of four-fermion operators. In the case of NC-NSI, the effective Lagrangian reads
\begin{align}
\mathcal{L_{\text{NC-NSI}}} = -2 \sqrt{2} G_F \varepsilon_{\alpha \beta}^{f X}\left(\overline{\nu}_{\alpha} \gamma^ \mu P_L \nu_{\beta}\right)\left(\overline{f} \gamma_{\mu} P_X f\right),
\label{eqn:lag}
\end{align}
where $G_F$ is the Fermi constant and the sum over the chirality of projectors ($X = \lbrace L,R \rbrace$), matter fields ($f = \lbrace
e, u, d \rbrace$), and flavours ($\alpha, \beta = \lbrace e, \mu, \tau \rbrace$) is implicit. The dimensionless coefficients $\varepsilon_{\alpha \beta}^{fX}$ quantify the strength of NSI with respect to SM interactions. Lepton flavour is not conserved in the presence of non-zero $\varepsilon_{\alpha\beta}^{fX}$ coefficients  with $\alpha \neq \beta$, whereas in the case of $\varepsilon_{\alpha\alpha}^{fX} - \varepsilon_{\beta\beta}^{fX} \neq 0$, NSI do not respect lepton flavour universality. Consequently, interactions are often classified into one of two categories: flavour changing NSI  and non-universal NSI, respectively.

In expression (\ref{eqn:lag}), the interactions can be projected onto the vector (V) and axial (A) components instead, so that $\varepsilon_{\alpha \beta}^{fV} = \varepsilon_{\alpha\beta}^{fL} + \varepsilon_{\alpha\beta}^{fR}$ and $\varepsilon_{\alpha \beta}^{fA} = \varepsilon_{\alpha\beta}^{fL} - \varepsilon_{\alpha\beta}^{fR}$. This parametrisation is particularly convenient when studying the impact of NSI on neutrino oscillations, since propagation is only affected by the vectorial component of interactions. Then, the Hamiltonian describing neutrino oscillations is  given by the sum of the vacuum Hamiltonian ($H_\text{vac}$) and the effective potentials due to both standard matter ($V_\text{SM}$) and NSI ($V_\text{NSI}$):
\begin{flalign}
\nonumber
H & =   H_\text{vac} + V_\text{SM} + V_\text{NSI} = \\
& =  U\frac{1}{2E}\begin{pmatrix} 
0 & 0 & 0 \\ 0 & \Delta m^2_{21} & 0 \\ 0 & 0 & \Delta m^2_{31}
\end{pmatrix} U^\dagger + \sqrt{2}G_F\left[N_e \begin{pmatrix}
1 & 0 & 0 \\ 0 & 0 & 0 \\ 0 & 0 & 0
\end{pmatrix} + \sum_{f = e,u,d}N_f \begin{pmatrix}
\varepsilon_{ee}^{fV} & \varepsilon_{e\mu}^{fV} & \varepsilon_{e\tau} ^{fV} \\ \varepsilon_{e\mu}^{fV*} & \varepsilon_{\mu\mu}^{fV} & \varepsilon_{\mu\tau}^{fV} \\ \varepsilon_{e\tau}^{fV*} & \varepsilon_{\mu\tau}^{fV*} & \varepsilon_{\tau\tau}^{fV}
\end{pmatrix}\right].
\label{eqn:ham}
\end{flalign}\\
Here, the lepton mixing matrix follows the usual parametrisation $U = U_{23}U_{13}U_{12}$, and  $N_f$ is the number density of the matter fields $f = \lbrace e, u, d \rbrace$ in the medium, which is assumed to be electrically neutral and unpolarised.

%%%%%%%%%%%%%%%%%%%%%%%%%%%%%%%%%%%%%%%%%%%%%
\subsection{Effective two-neutrino approach}
%%%%%%%%%%%%%%%%%%%%%%%%%%%%%%%%%%%%%%%%%%%%%

In the absence of NSI, the evolution of solar neutrinos within the Sun and through the Earth satisfies the condition $\sqrt{2}G_F N_e \lesssim \Delta m^2_{21}/2E \ll |\Delta m^2_{31}|/2E$, with $\theta_{13} \ll 1$. This means neutrino oscillations can be studied using an effective two-neutrino approach, where the evolution of a third eigenstate decouples from the other two. 
This also applies to the evolution of solar neutrinos in the presence of NSI, where the effective potential $V_\text{NSI}$ is of the same order as the standard effective potential in matter, $V_\text{SM}$.
Under this approximation, the survival probability for electron neutrinos is given by~\cite{Lim:1987yd,Kuo:1989qe}
\begin{flalign}
P_{ee} = \cos^4\theta_{13} P^{2\nu}_{ee} + \sin^4 \theta_{13},
\end{flalign}
where the effective survival probability $P^{2\nu}_{ee}$ is calculated from the two-neutrino effective Hamiltonian 
\begin{flalign}
H^{2\nu} = \frac{\Delta m^2_{21}}{4E}\begin{pmatrix}
-\cos 2 \theta_{12} & \sin 2\theta_{12} \\ \sin 2 \theta_{12}  & \cos 2\theta_{12}
\end{pmatrix} + \sqrt{2}G_F \left[ \cos^2\theta_{13}N_e \begin{pmatrix}
1 & 0 \\ 0 & 0
\end{pmatrix} +  \sum_{f = e, u, d} N_f\begin{pmatrix}
0 & \varepsilon_f \\ \varepsilon^{*} _f & \varepsilon' _f
\end{pmatrix}\right],
\label{eqn:eff}
\end{flalign}
describing the evolution of the state $\nu = ( \nu_e, \nu_x)^T$, with $\nu_x$ being a mixture of $\nu_\mu$ and $\nu_\tau$, in the presence of NSI. 
The effective NSI parameters $\varepsilon_f$ and $\varepsilon ' _f$ account for flavour-changing and non-universal NSI, respectively. They are related to the NSI parameters introduced in Eqs.~(\ref{eqn:lag}) and (\ref{eqn:ham}) in the following way~\cite{Friedland:2004pp, Miranda:2004nb,Escrihuela:2009up,Esteban:2018ppq}: 
\begin{align}
\begin{aligned}
    \varepsilon_f =  \sin \theta_{13} e^{-i \delta_{CP}}\left[\sin^2\theta_{23} \varepsilon^{fV}_{\mu \tau} - \cos^2 \theta_{23} \varepsilon_{\mu \tau }^{*fV} + \left( \varepsilon^{fV}_{\tau \tau} - \varepsilon^{fV} _{\mu \mu}\right)\cos \theta_{23} \sin \theta_{23}\right] \\ + \cos \theta_{13} \left(\cos \theta_{23} \varepsilon^{fV}_{e\mu} - \sin\theta_{23} \varepsilon^{fV}_{e\tau}\right)
\end{aligned}
\label{eqn:eps}
\end{align}
and
\begin{align}
    \begin{aligned}
        \varepsilon_f ' = 2 \cos \theta_{13} \sin \theta_{13} \text{Re} \left[e ^ {i \delta_{CP}} \left(\cos \theta_{23} \varepsilon^{fV}_{e \tau} + \sin\theta_{23} \varepsilon^{fV}_{e\mu}\right)\right] - 2\left(1 + \sin ^2\theta_{13}\right) \cos \theta_{23} \sin\theta_{23} \text{Re}\left[\varepsilon^{fV}_{\mu \tau}\right] \\ + \varepsilon^{fV}_{\mu\mu} \left(\cos^2 \theta_{23} - \sin ^2 \theta_{13}  \sin ^2 \theta_{23}\right) + \varepsilon^{fV}_{\tau\tau} \left(\sin^2 \theta_{23} - \sin^2\theta_{13} \cos^2\theta_{23}\right) - \cos^2\theta_{13} \varepsilon^{fV}_{ee}.
    \end{aligned}
\label{eqn:epsprime}
\end{align}

This effective two-neutrino description is also valid for medium-baseline reactor experiments if their energy resolution  is not good enough to resolve the subleading oscillation interference between $\Delta m^2_{31}$ and $\Delta m^2_{32}$. Likewise, it applies to long-baseline reactor experiments, which are not sensitive to these mass splittings due to the long neutrino flight paths involved. Although it is not possible to understand the physics expected in a medium-baseline reactor experiment like JUNO using an effective two-neutrino framework, we will see that it can nevertheless be useful in order to gain a better understanding of the impact of non-standard interactions. It should be noted, however, that although some results will be presented in the two-neutrino approximation for illustrative purposes, a three-neutrino numerical approach was followed throughout the analysis.

%%%%%%%%%%%%%%%%%%%%%%%%%%%%%%%%%%%%%%%%%%%%%%%%%%%%%%%%%%%%%%%%%%%%%%%%%
\subsection{Generalised mass ordering degeneracy and the LMA-D solution}
%%%%%%%%%%%%%%%%%%%%%%%%%%%%%%%%%%%%%%%%%%%%%%%%%%%%%%%%%%%%%%%%%%%%%%%%%

The evolution of a three-flavour system, as described by the Hamiltonian for propagation in vacuum $H_{vac}$ in Eq.~(\ref{eqn:ham}), remains invariant under the transformation
\begin{flalign}
\label{eqn:paramsdark}
\theta_{12} \longrightarrow \pi/2 - \theta_{12}, \indent \Delta m^2_{31} \longrightarrow - \Delta m^2 _{31} + \Delta m^2 _{21} = - \Delta m^2_{32} \indent \text{and} \indent \delta_{CP} \indent \longrightarrow \indent \pi - \delta_{CP}, 
\end{flalign}
which gives rise to the well-known generalised mass ordering degeneracy in vacuum~\cite{Coloma:2016gei}. Matter effects break this degeneracy and, consequently, solar neutrino experiments can determine $\sin^2 \theta_{12} < 0.5$. Notwithstanding, the degeneracy can be recovered in the presence of NSI when NSI parameters transform as follows:
\begin{flalign}
\label{eqn:epsdark}
\sum_{f = e,u,d} N_f \varepsilon_{ee}^{fV} \longrightarrow - \sum_{f = e,u,d} N_f \varepsilon_{ee}^{fV} - 2N_e \indent \text{and} \indent \sum_{f = e,u,d} N_f \varepsilon_{\alpha \beta}^{fV} \longrightarrow - \sum_{f = e,u,d} N_f \varepsilon_{\alpha \beta}^{fV*} (\alpha\beta \neq ee)\, .
\end{flalign}
As long as the mass ordering remains undetermined through means that are not affected by NSI, the degeneracy in the Hamiltonian set by the transformations in Eqs.~(\ref{eqn:paramsdark}-\ref{eqn:epsdark}) cannot be easily resolved. Then, a solution with $\sin^2 \theta_{12} > 0.5$ becomes possible; this is known as the LMA-D solution  of the solar neutrino problem~\cite{Miranda:2004nb,Escrihuela:2009up}. 

This result can be translated into the two-neutrino approach, described by the Hamiltonian in (\ref{eqn:eff}). In this case, the evolution of the two-flavour system remains invariant under the transformation 
\begin{flalign}
\label{eqn:epsdark2}
\theta_{12} \longrightarrow \pi/2 - \theta_{12}, \nonumber \\ \sum_{f = e,u,d} N_f \varepsilon_f ' \longleftrightarrow - \sum_{f = e, u,d}N_f \varepsilon_f ' + 2\cos^2 \theta_{13}N_e \indent \text{and} \indent \sum_{f = e,u,d}N_f \varepsilon_f \longrightarrow -\sum_{f = e, u,d}N_f \varepsilon_f ^*.
\end{flalign}
It should be noted that the LMA-D solution is only possible for very large diagonal NSI parameters. Scattering experiments, including coherent elastic neutrino-nucleus scattering, can severely constrain NSI and thus help to resolve the degeneracy \cite{Escrihuela:2009up,Esteban:2018ppq, EstevesChaves:2021jct}. In addition, the combination of solar and long or medium-baseline reactor experiments can also slightly lift the degeneracy due to the fact that the number density of the matter fields $N_f$ is neither constant nor equal on Earth and in the Sun. Nonetheless, reactor experiments are not very sensitive to matter effects, which is the main limitation for this approach.

%%%%%%%%%%%%%%%%%%%%%%%%%%%%%%%%%%%%%%%%%%%%%%
%%%%%%%%%%%%%%%%%%%%%%%%%%%%%%%%%%%%%%%%%%%%%%
\section{Hyper-Kamiokande}
\label{sec:hk}
%%%%%%%%%%%%%%%%%%%%%%%%%%%%%%%%%%%%%%%%%%%%%%
%%%%%%%%%%%%%%%%%%%%%%%%%%%%%%%%%%%%%%%%%%%%%%
\subsection{Simulation and analysis}
%%%%%%%%%%%%%%%%%%%%%%%%%%%%%%%%%%%%%%%%%%%%%%

Following the success of the Kamiokande and Super-Kamiokande experiments, Hyper-Kamiokande~\cite{Abe:2018uyc} will be the next-generation water Cherenkov detector in Japan. With a fiducial volume of 187\,kton, 8.3 times greater than that of Super-Kamiokande (and greater still if two tanks are built), Hyper-Kamiokande will have a huge multipurpose research potential. It will be capable of studying everything from solar and atmospheric neutrinos to supernovae, as well as having applications to dark matter searches and neutrino tomography. Its three principal physics goals revolve around CP violation, neutrino mass ordering and nucleon decay, and the long-baseline aspect of the collaboration, T2HK, will form part of the next generation of oscillation experiments. In this work, however, we will focus on its expected capacity to measure and study solar neutrinos.

In order to estimate the projected sensitivity for the Hyper-Kamiokande experiment, we studied three different possible configurations based on its potential fiducial volume and low-energy threshold. These are summarised in Table \ref{tab:confs}: in the first configuration, we assume an energy threshold lower than that currently planned by the collaboration (but consistent with Super-Kamiokande's most recent achievements), while the second one is a more conservative estimate based on an expected lower overall photocoverage compared to Super-Kamiokande's 40\,\%. The third configuration takes into account the possibility of a second tank, located either in Japan or South Korea, being built at a later date~\cite{Abe:2018uyc}. We consider 10 years of runtime in every case except for in Configuration C, where we assume the second tank will operate for a further 3 years.\\

\begin{table}[!t]
    \begin{tabular}{lccc} \toprule
	     Configuration \, & Energy threshold (MeV) \, & Fiducial volume (kton) \, & Running time (years) \, \\
	    \colrule
	     A (optimistic) & 3.5 & 187 & 10 \\
	     B (conservative) & 5 & 187 & 10 \\
	     C (2 tanks) & 5 & 187 & $10 + 3$ \\ \botrule
        \end{tabular}
    \caption{
        \label{tab:confs}
        Main characteristics of the three possible configurations studied for Hyper-Kamiokande.
    }
\end{table}
A light water detector such as Hyper-Kamiokande is sensitive  to solar neutrinos only through neutrino-electron elastic scattering: $\nu_x + e^- \rightarrow \nu_x + e^-$ . For our simulation, the corresponding cross-section was taken from \cite{Bahcall:1995mm} and the response of the detector was estimated using a Gaussian function with the same energy resolution as in Super-Kamiokande Run IV (SK-IV) \cite{Abe:2016nxk}.

Following the analysis in Ref.~\cite{Nakano:2015wdv}, one can define an extended $\chi^2$ function which includes spectral as well as zenithal information in the form of day and night energy bins. %
Instead of computing the absolute number of events in every bin, we will perform the analysis in terms of the ratio between the number of events with and without flavour oscillations,
\begin{align}
    r_{i,j,k} = \frac{r_{i,j,k}^{\text{ osc}}}{r_{i,^8B, k}^{\text{ unosc}}+r_{i,\text{hep}, k}^{\text{ unosc}}}\, .
\end{align}
The index $i$ indicates the energy bin, $j\in  \lbrace ^8B, hep \rbrace$ indicates the source of the neutrino flux, and $k$ refers to the zenith angle binning, labelled as day (D) or night (N). 
Our $\chi^2$-function thus reads:
\begin{eqnarray}
    \chi^2 & = & \sum_{k = D,N}\sum_{i = 1}^{i = 23} \frac{\left[d_{i,k} - b_{i,k} ( \alpha, \epsilon_{^8B}, \epsilon_{scale}, \epsilon_{resol}) - h_{i,k}( \beta)\right]^2}{(\sigma^{i,k}_{stat})^2 + (\sigma^i_{uncorr})^2}  
    + \left(\frac{\alpha}{\sigma_{\alpha}}\right)^2 + \left(\frac{\beta}{\sigma_{\beta}}\right)^2 + \nonumber \\ 
    & + & \epsilon^2_{^8B}+ \epsilon^2_{scale} + \epsilon^2_{resol} \, ,
\end{eqnarray}
where we have defined
\begin{align}
      \indent b_{i,j}&(\alpha, \varepsilon_{^8B}, \epsilon_{scale}, \epsilon_{resol}) =  (1 + \alpha + \epsilon_{^8B} \cdot \sigma ^{i,k}_{^8B} + \epsilon_{scale}\cdot \sigma^{i,k}_{scale} + \epsilon_{resol}\cdot \sigma^{i,k}_{resol}) \cdot r_{i, ^8 B, k} \, , &&\\
      \indent h_{i,k}&(\beta)  =  (1 + \beta)\cdot r_{i, hep, k} \, .&&
\end{align}

The complete function depends on the ``observed" number of events per energy bin $i$ and zenith bin $k \in \lbrace D,N \rbrace$, $d_{i,k}$, generated as mock data assuming the best fit value for the neutrino oscillation parameters from \cite{deSalas:2020pgw} (see Table \ref{tab:oscparams}). The theoretically estimated number of events from the $^8B$ chain, $b_{i,k}$, includes contributions from energy-correlated systematics due to the flux shape uncertainty $\sigma^{i,k}_{^8B}$, the energy scale $\sigma^{i,k}_{scale}$, and the energy resolution $\sigma^{i,k}_{resol}$\footnote{It should be noted that the dependence of $b_{i,j}$, $h_{i,j}$, $\sigma^{i,k}_{^8B}$, $\sigma^{i,k}_{scale}$ and $\sigma^{i,k}_{resol}$ on the oscillation parameters is not indicated explicitly.}.
Such contributions are not included in the prediction from the $hep$ chain, as this flux already provides a subdominant contribution to the signal. These energy-correlated uncertainties are weighted by three corresponding nuisance parameters ($\epsilon_{^8B}$, $\epsilon_{scale}$ and $\epsilon_{resol}$). Two additional nuisance parameters, $\alpha$ and $\beta$, are included in order to account for uncertainties in the total normalization of $^8B$ and $hep$ solar neutrino fluxes, respectively. The corresponding penalty terms
as well as the  statistical and energy-uncorrelated uncertainties,  
$\sigma^{i,k}_{stat}$ and  $\sigma^i_{uncorr}$,  
are also included in the $\chi^2$ function. 
Energy-uncorrelated systematics were assumed to be equal to those in SK-IV, as in Table 10.2 from Ref.~\cite{Nakano:2015wdv}, while statistical systematics were scaled from those of Super-Kamiokande, assuming that the number of events follows a Poissonian distribution. In this case, the standard deviation will be given by $\sigma^i_{stat} = \sqrt{N^i_{events}}$ for each bin $i$. For an updated analysis with a longer running time, $T_{HK}/ T_{SK} > 1$, and a larger volume, $V_{HK}/V_{SK} >1$, one will have reduced statistical errors:
\begin{align}
    \frac{\sigma^i_{stat, HK}}{N^i_{events,SK}} = \frac{1}{\sqrt{N^i_{events, HK}}} = \sqrt{\frac{1}{N^i_{events, SK}}\frac{T_{SK}}{T_{HK}} \frac{V_{SK}}{V_{HK}}} = \frac{\sigma^i_{stat, SK}}{N^i_{events,SK}} \sqrt{\frac{T_{SK}}{T_{HK}} \frac{V_{SK}}{V_{HK}}}.
\end{align}

Regarding uncertainties on the total flux, $\sigma_\alpha = 4\%$ is taken from the NC measurement carried out by the SNO collaboration \cite{Aharmim:2009gd}, while $\sigma_\beta = 200\%$ is a conservative choice as in Ref.~\cite{Nakano:2015wdv}.

\begin{table}[tb]
       \begin{tabular}{cc} \toprule
   \multicolumn{2}{c} {Neutrino oscillation parameters} \\
   \colrule
	       $\sin^2 \theta_{12} = 0.32$ \hphantom{mmmm} & $\Delta m^2_{21} = 7.5~\times 10^{-5}$ eV$^2$ \quad \\ 
	       $\sin^2 \theta_{13} = 0.022$  \hphantom{mmmm} &  $\Delta m^2_{31} = 2.55 \times 10 ^{-3}$ eV$^2$ \quad   \\ 
	       $\sin^2 \theta_{23} = 0.574$  \hphantom{mmmm} &  $\delta_{CP} = 1.2 \pi$    \quad \\  \botrule
        \end{tabular}
    \caption{
        \label{tab:oscparams}
        Best fit values for the oscillation parameters, as from \cite{deSalas:2020pgw}.
    }
\end{table}

%%%%%%%%%%%%%%%%%%%%%%%%%%%%%%%%%%%%%%%%%%%%%%%%%%%%%%%%%%%%%
\subsection{Impact of NSI on solar neutrino experiments}
%%%%%%%%%%%%%%%%%%%%%%%%%%%%%%%%%%%%%%%%%%%%%%%%%%%%%%%%%%%%%
As stated above, even after the inclusion of a new interaction framework, the evolution of neutrinos inside the Sun remains adiabatic. However, NSI alter the neutrino mixing in the production region, which translates into a change in the position and shape of the transition region in the energy profile. Neglecting matter effects on $\sin^2 \theta_{13}$, which are known to be small \cite{Li:2016txk, An:2015jdp}, the survival probability during the day is given by
\begin{flalign}
P^{D}_{ee,\odot} = \cos ^4\theta_{13} \left[\cos^2\theta_{13} \cos^2 \tilde{\theta}_{12} + \sin^2\theta_{12} \sin^2\tilde{\theta}_{12}\right] + \sin^4 \theta_{13},
\end{flalign}
where we have defined the mixing at the production point in the Sun as
\begin{flalign}
\cos{2\tilde{\theta}_{12}} = \frac{\Delta m^2_{21}\cos 2\theta_{12}+ 2 \sqrt{2}G_FE \left(\sum_{f= e,u,d} N^0_f \varepsilon ' _f - N^0_e\right)}{\Delta \tilde{m}^2_{21}},
\label{eqn:angle}
\end{flalign}
with
\begin{eqnarray}
\left[\Delta \tilde{m}^2_{21}\right]^2 & = & \left[\Delta m^2_{21}\cos 2\theta_{12} + 2 \sqrt{2}G_F E \left(\sum_{f= e,u,d} N^0_f \varepsilon ' _f - N^0_e\right) \right]^2 \nonumber \\& + & \left[ \Delta m^ 2_{21} \sin 2 \theta_{12} + 4\sqrt{2} G_F E \sum_{f = e,u,d}N^0_f \varepsilon_f \right]^2 \,. 
\label{eqn:mass}
\end{eqnarray}
In the above expressions, $N^0_f$ refers to the number density of the matter fields $f = e, u,d$ in the solar neutrino production region.

Solar neutrinos can also travel through the Earth before reaching the detector. In this scenario, it should be taken into account that, in spite of arriving from the Sun as an incoherent admixture of mass eigenstates, they will undergo flavour oscillations as they traverse the Earth. A zenith-angle dependence then arises in the oscillation probability, which is generally referred to as the day-night asymmetry. This observable is often computed numerically, since it requires solving the evolution of the system in a varying matter potential; nevertheless, a good understanding can be gained through careful analytical studies~\cite{Blennow:2003xw,Akhmedov:2004rq,Ioannisian:2004vv}. In the presence of NSI, there are also analytical expressions which can help understand the overall picture \cite{Friedland:2004pp, Liao:2017awz}. 

\begin{figure}[b!]
\centering
\includegraphics[width=0.6\textwidth]{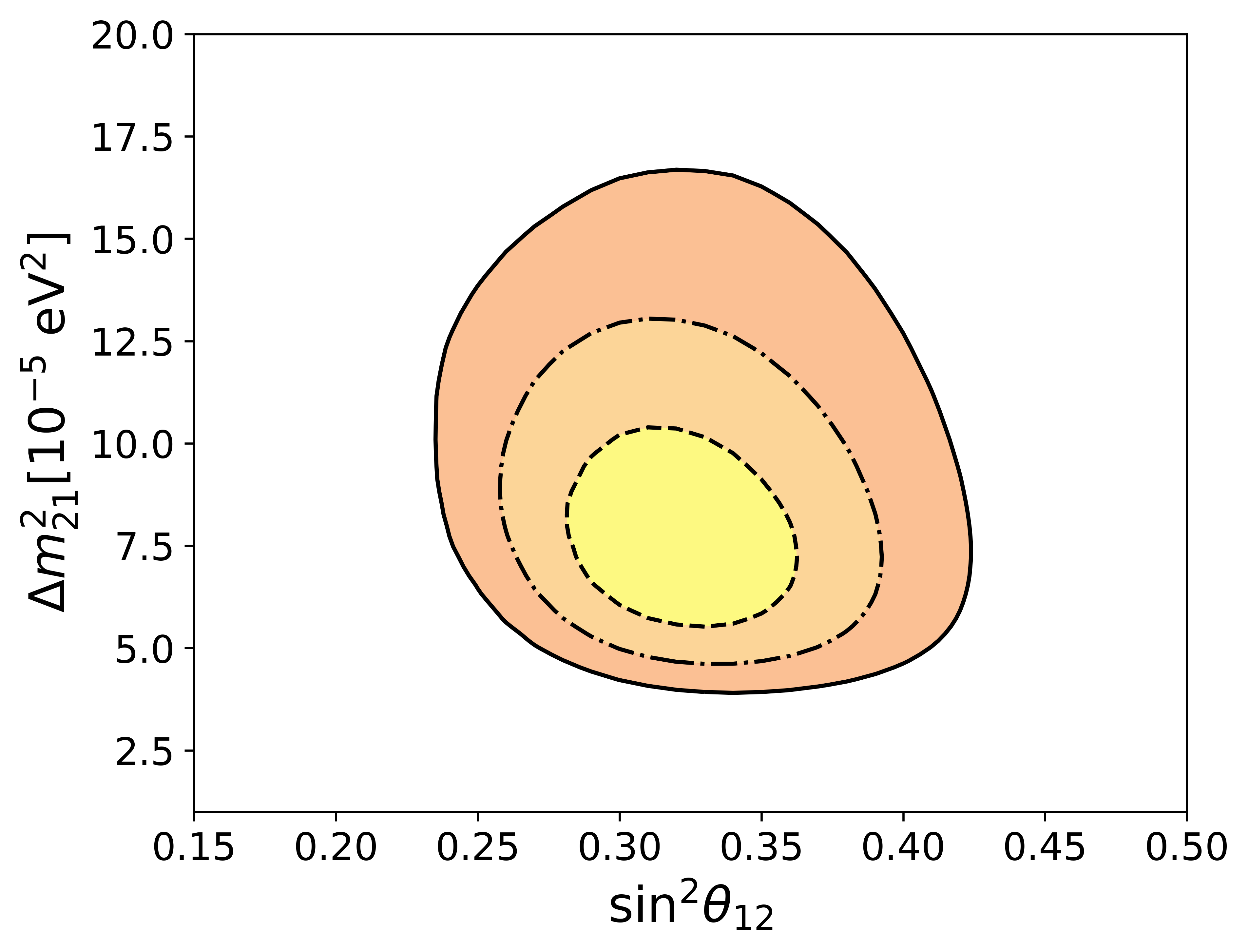}
\caption{
\label{fig:stdHK}
Expected sensitivity to the solar neutrino oscillation parameters in Hyper-Kamiokande in the absence of NSI, where 1$\sigma$, 2$\sigma$ and 3$\sigma$ confidence levels are indicated by the dashed, dot-dashed and solid lines, respectively. }
\end{figure}

As a consequence of both the different mixing in the production region and the modified propagation on Earth, one would expect NSI to greatly distort the standard picture, inducing potentially large shifts in the values of oscillation parameters. As a reference point, Figure \ref{fig:stdHK} shows the expected sensitivity of Hyper-Kamiokande to solar oscillation parameters in the absence of NSI, assuming 10 years of running time and a 3.5 MeV threshold, referred to as Configuration A in Table \ref{tab:confs}. 

%%%%%%%%%%%%%%%%%%%%%%%%%%%%%%%%%%%%%%%%%%%%%%
%%%%%%%%%%%%%%%%%%%%%%%%%%%%%%%%%%%%%%%%%%%%%%
\section{JUNO}
%%%%%%%%%%%%%%%%%%%%%%%%%%%%%%%%%%%%%%%%%%%%%%
%%%%%%%%%%%%%%%%%%%%%%%%%%%%%%%%%%%%%%%%%%%%%%
\label{sec:juno}
\subsection{Simulation and analysis}
%%%%%%%%%%%%%%%%%%%%%%%%%%%%%%%%%%%%%%%%%%%%%%

The Jiangmen Underground Neutrino Observatory (JUNO) \cite{Djurcic:2015vqa} is a next-generation medium-baseline reactor experiment. In the same way as KamLAND~\cite{KamLAND:2008dgz}, JUNO will detect reactor neutrinos through inverse beta decay (IBD), as it will be sensitive to the disappearance of electron antineutrinos. Thanks to its expected increase in statistics and improved energy resolution with respect to those of \mbox{KamLAND}, JUNO will yield significant advantages when it comes to performing precision measurements. It will have a fiducial volume of 20\,kton of liquid scintillator (20 times larger than that of KamLAND) and an average baseline of 53\,km compared to KamLAND's 180\,km.

The main contributions to the antineutrino flux in JUNO will come from the Yangjiang and Taishan Nuclear Power Plants, located approximately 53\,km away from the detector. The first power plant consists of 6 cores with a thermal power of 2.9\,GW, while the second one has 2 cores with 4.6\,GW of power each. In addition, the Daya Bay and Huizhou complexes will give a non-negligible contribution to the neutrino signal expected. In our analysis, we have treated these last two power plants as two cores located at baselines of 215\,km and 265\,km respectively, meaning we considered the total antineutrino flux to have contributions from 12 reactor cores.

The energy resolution expected at JUNO is $3\,\% /\sqrt{E(\text{MeV})} $. This will allow a precise measurement of the solar oscillation parameters $\theta_{12}$ and $\Delta m^2_{21}$, as well as a determination of the mass ordering \cite{An:2015jdp}.
Inspired by the oscillation analyses in Refs.~\cite{An:2015jdp,Bezerra:2019dao}, we use 200 equal-size bins for the incoming neutrino energy ranging from 1.8\,MeV to 8.0\,MeV and define the following $\chi^2$ function:
\begin{flalign}
\chi^2 = \sum_{i = 1}^{200} \frac{\left[N_i - \sum_{j = 1}^{12} (1 + \xi _a)(1+ \xi_{r,j})(1+\xi_{s,i})T_{ij}\right]^2}{N_i (1+ \sigma_d N_i)} + \sum_{j=1}^{10} \left(\frac{\xi_{r,j}}{\sigma_r}\right)^2 + \left( \frac{\xi_a}{\sigma_a}\right)^2 + \sum_{i=1}^{200} \left(\frac{\xi_{s,i}}{\sigma_{s}}\right), 
\end{flalign}
where $N_i$ denotes the observed number of events in the $i$-th energy bin that we simulate as being the expected ones from the best fit in \cite{deSalas:2020pgw} and $T_{ij}$  refers to the predicted number of events in the $i$-th energy bin due to the $j$-th reactor core for the set of parameters that is being tested. 
Regarding systematic uncertainties, they are accounted for by introducing a total of  nuisance  211 nuisance parameters. We have included an absolute uncertainty on the reactor flux, $\sigma_a=2\,\%$ (the associated nuisance parameter is denoted by $\xi_{a}$), an  uncertainty related to each reactor, $\sigma_{r} = 0.08\,\%$ (the corresponding pull parameters are $\xi_{r,j}$, with $j \in \lbrace 1, 10\rbrace$), and an  uncertainty on the shape of the spectrum, $\sigma_{s}=1\,\%$ (the pull parameters included for the $i$-th bin are $\xi_{s,i}$, with $i \in \lbrace 1,200 \rbrace$). An uncorrelated uncertainty from the detector, $\sigma_{d} = 1\,\%$, is also included. In our calculations, we implement the IBD cross-section as in \cite{Vogel:1999zy}, the energy spectra from \cite{Mueller:2011nm}, and the reactor fission fractions from \cite{Zhan:2008id}. Event computation and the minimisation of our $\chi^2$ function were performed using GLoBES (General Long Baseline Experiment Simulator) \cite{Huber:2004ka, Huber:2007ji}.

\begin{figure}[b!]
\centering
\includegraphics[width=0.6\textwidth]{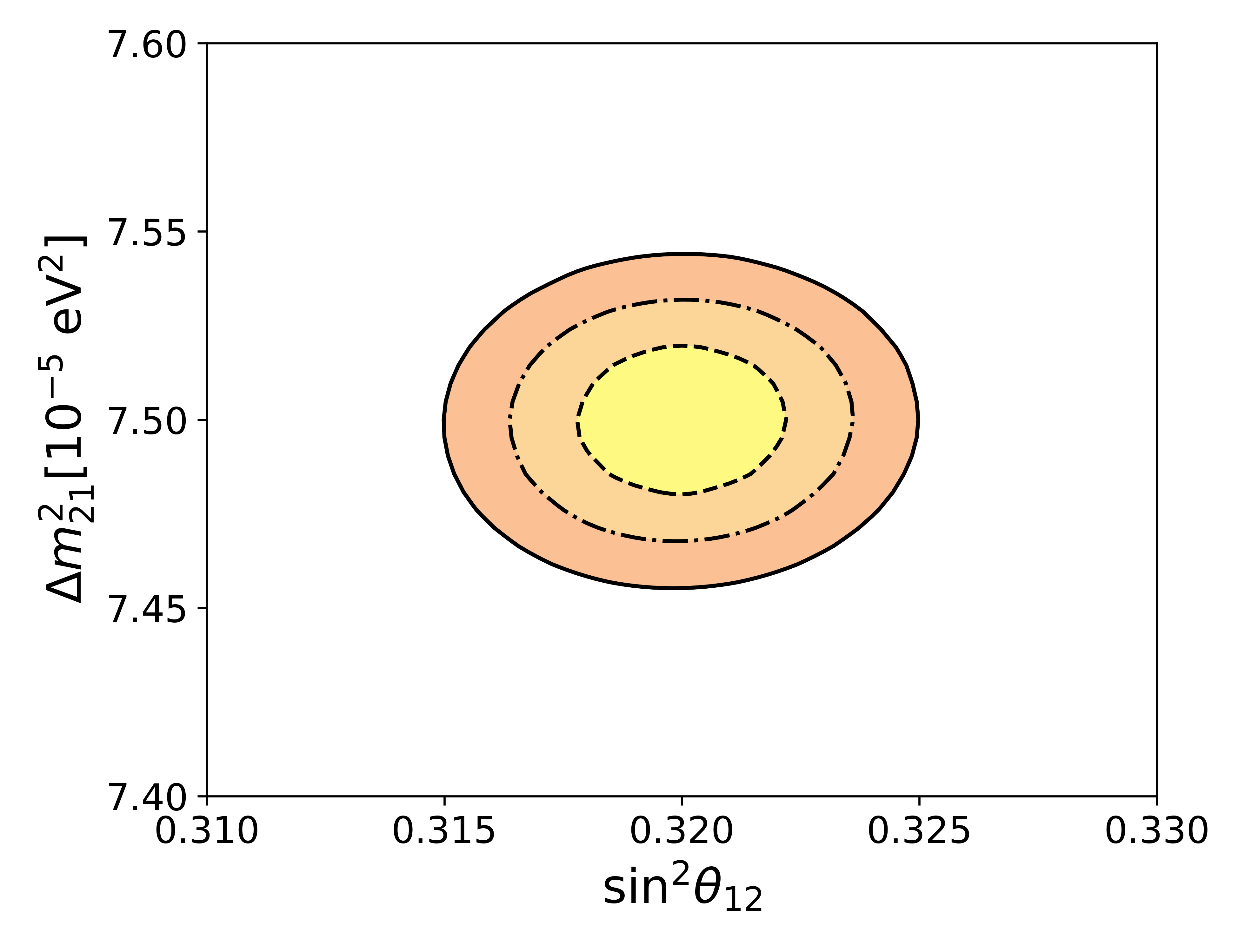}
\caption{
\label{fig:stdJUNO}
 JUNO's expected sensitivity to the oscillation parameters $\theta_{12}$ and $\Delta m^2_{21}$ in the absence of NSI, where the 1$\sigma$, 2$\sigma$ and 3$\sigma$ C.L. contours are shown as before.}
\end{figure}
%
%%%%%%%%%%%%%%%%%%%%%%%%%%%%%%%%%%%%%%%%%%%%%%%%%%%%%%%%%%%%%%%%%%
\subsection{Impact of NSI on medium-baseline reactor antineutrino experiments}
%%%%%%%%%%%%%%%%%%%%%%%%%%%%%%%%%%%%%%%%%%%%%%%%%%%%%%%%%%%%%%%%%%
If matter effects and NSI are not considered, the survival probability in medium-baseline reactor experiments is given by:
\begin{flalign}
P^{\rm MBL-Reac}_{\overline{e}\overline{e}} = 1- \cos^4\theta_{13}\sin^2 2\theta_{12}\sin^2\Delta_{21} - \sin^2 2\theta_{13}\left(\cos ^2\theta_{12}\sin ^2 \Delta_{31} + \sin^2 \theta_{12}\sin^2 \Delta_{32}\right) ,
\label{eqn:junop}
\end{flalign}
where we have defined $\Delta_{ij} = \Delta m^2 _{ij} L /4E$. 

 Non-standard neutrino interactions with matter will have a similar impact on oscillation parameters in the solar sector as those discussed in Eqs.~(\ref{eqn:angle}) and (\ref{eqn:mass}) for solar neutrinos, with the key difference that, since reactors emit electron antineutrinos, the matter and NSI terms in the Hamiltonian will have an opposite sign to their counterparts in the case of neutrinos. For completeness, Figure \ref{fig:stdJUNO} shows the expected sensitivity of JUNO to $\sin^2\theta_{12}$ and $\Delta m^2_{21}$ in the absence of NSI, after marginalising over $\theta_{13}$ and $\Delta m^{2}_{31}$ for normal ordering. 

With respect to the other two oscillation parameters influencing the survival probability expected at JUNO, $\theta_{13}$ and $\Delta m^2_{31}$, they are not significantly affected by matter effects or NSI according to current constraints \cite{An:2015jdp}. Hence, one would expect JUNO to be capable of providing an accurate measurement of these oscillation parameters even in the presence of NSI.

%%%%%%%%%%%%%%%%%%%%%%%%%%%%%
%%%%%%%%%%%%%%%%%%%%%%%%%%%%%
\section{Results}
\label{sec:results}
%%%%%%%%%%%%%%%%%%%%%%%%%%%%%
%%%%%%%%%%%%%%%%%%%%%%%%%%%%%
\subsection{NSI in Hyper-Kamiokande}
\label{sec:NSI-HK}

Neutrino non-standard interactions are known to alter the solar neutrino picture considerably, since they affect the mixing in the production region of the Sun and modify the day-night asymmetry expected from neutrino propagation through the Earth. In addition, for the experimental configuration we are considering here, only one side of the neutrino spectrum is accessible, with energies above a certain threshold. This means that, although the large number of statistics expected would allow a differentiation between the spectra for day and night, the transition region and the low-energy side of the neutrino spectrum will not be measurable and Hyper-Kamiokande will have to rely on previous measurements from other experiments.

At this point, it is important to remember that we are restricting ourselves to the case of NSI with $d$-type quarks. We also introduce here the short-hand notation $\varepsilon ' _f = \varepsilon '$ and $\varepsilon _f = \varepsilon$ for $f = d$.

\begin{figure}[b!]
\centering
\includegraphics[width=\textwidth]{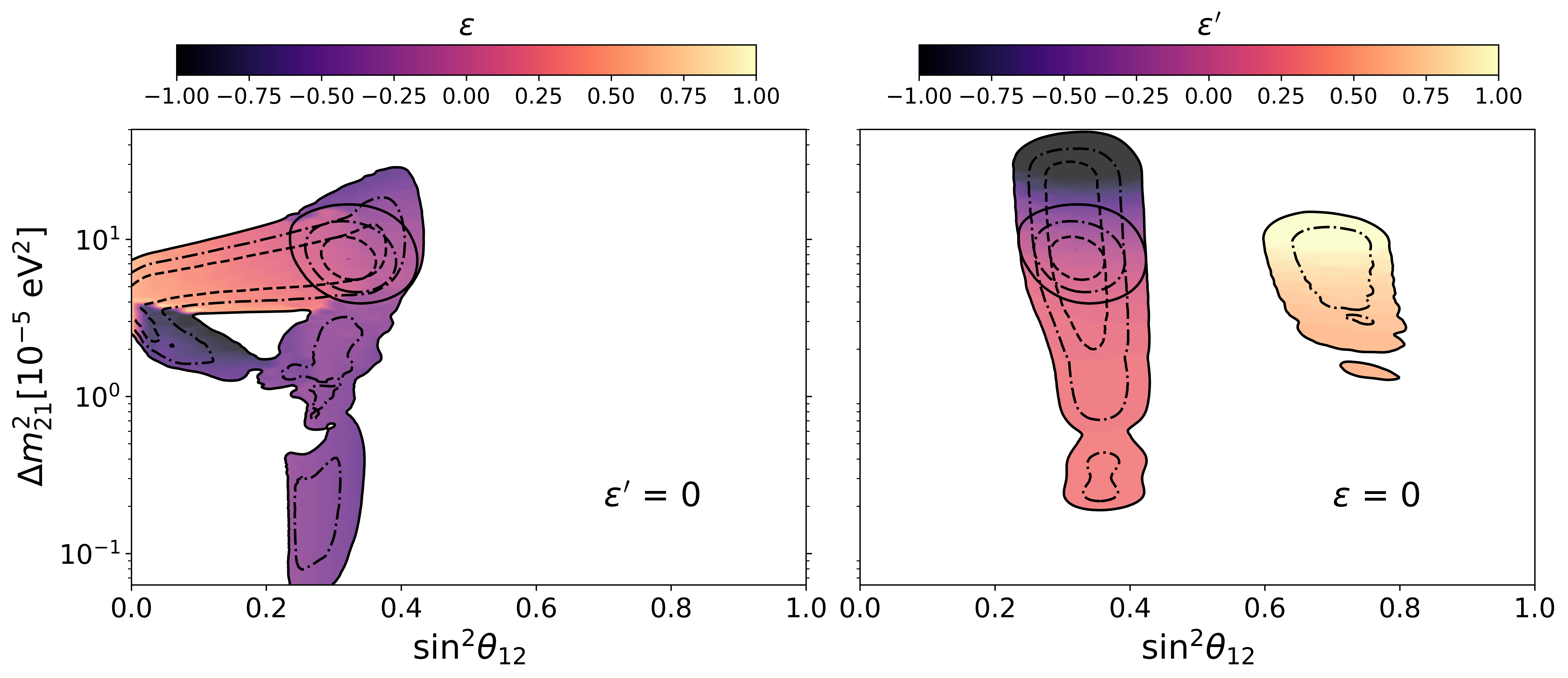}
\caption{
\label{fig:oneatatimeHK}
Effect of the effective NSI parameters on the sensitivity of Hyper-Kamiokande, varying one at a time. Left panel shows the 1$\sigma$, 2$\sigma$ and 3$\sigma$ C.L. in the $\sin^2 \theta_{12}- \Delta m^2_{21}$ plane when varying $\varepsilon$ between -1 and 1. Same confidence levels are drawn in the right panel for the case of $\varepsilon = 0$ and $\varepsilon '$ allowed to vary within the same range. The colour map indicates the best fit value of the effective NSI parameters. Unfilled contours correspond to the same confidence levels expected for Hyper-Kamiokande in the absence of NSI. }
\end{figure}

In the right panel of Figure \ref{fig:oneatatimeHK}, we show the projected sensitivity of Hyper-Kamiokande to solar oscillation parameters in the presence of a non-zero NSI coefficient, $\varepsilon'$. It can be seen that non-universal NSI would affect the determination of the mass splitting by more than an order of magnitude. Besides this, it should be noted that a solution in the second octant arises for very large values of $\varepsilon '$; this corresponds to the LMA-D solution discussed in Section~\ref{sec:teo}~\cite{Miranda:2004nb}. Both of these features are expected from the arguments presented in previous sections.
Regarding flavour-changing NSI, there is a strong degeneration of the effective parameter $\varepsilon$ with the oscillation parameters $\sin^2\theta_{12}$ and $\Delta m^2_{21}$. This is shown in the left panel of Figure \ref{fig:oneatatimeHK}, where it can be seen how the allowed parameter space in this plane is significantly enlarged with respect to the standard LMA solution in the absence of NSI~\cite{deSalas:2020pgw}.

\begin{figure}[h!]
\centering
\includegraphics[width=\textwidth]{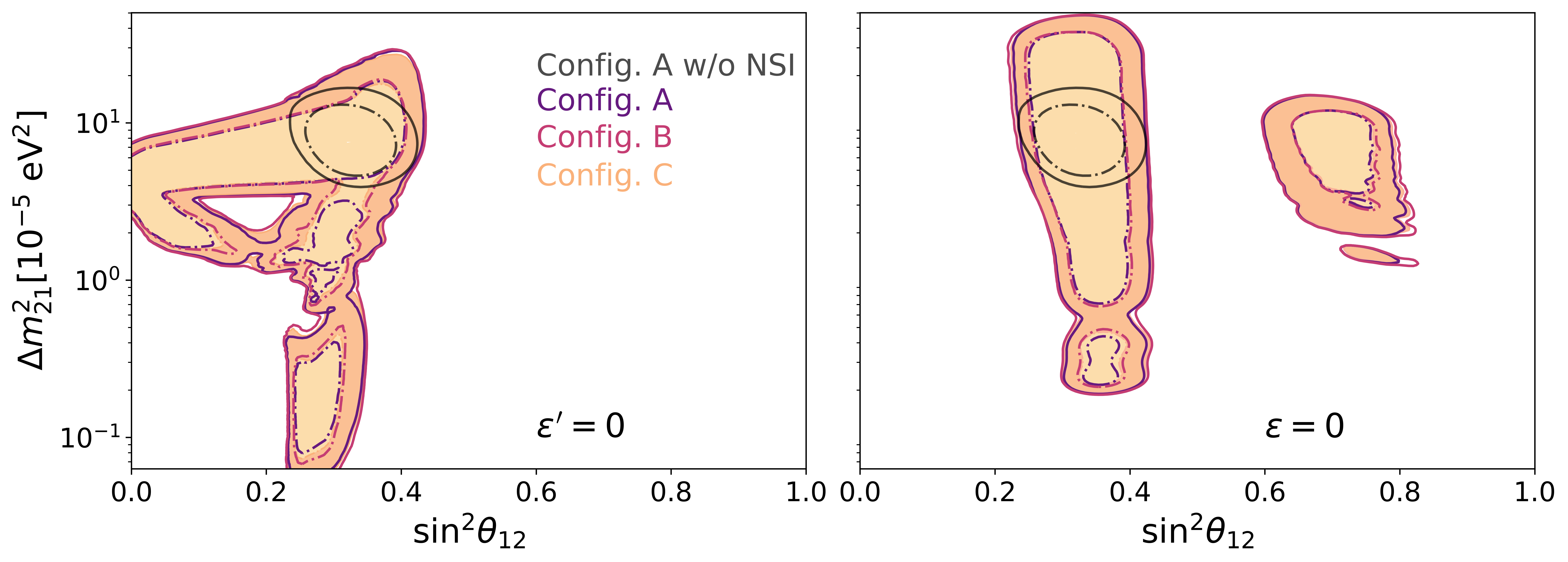}
\caption{
\label{fig:HK-confs}
 Effect of effective NSI parameters on the sensitivity of Hyper-Kamiokande, varying one at a time, for each of the three experimental configurations in Table~\ref{tab:confs}. Contours correspond to 2$\sigma$ and 3$\sigma$ C.L. In the left panel, $\varepsilon'$ is fixed to zero; in the right panel, $\varepsilon = 0$ is considered. Unfilled contours correspond to the same confidence levels expected for Hyper-Kamiokande in the absence of NSI for Configuration A.}
\end{figure}

In order to break these degeneracies, which increase substantially when both non-universal and flavour-changing NSI are considered simultaneously, the inclusion of other datasets is crucial, as Hyper-Kamiokande cannot resolve them by measuring the high-energy range of solar neutrinos alone. In fact, though small differences in this energy range are expected in the presence of NSI, an experimental configuration aimed at maximising statistics, i.e. one involving two tanks, would not be able to set significant constraints on NSI parameters on its own. Similarly, lowering the energy threshold to $3.5$\,MeV would not help to resolve the degeneracies of $\varepsilon$ and $\varepsilon '$ with the oscillation parameters.
This can be seen in Figure \ref{fig:HK-confs}, where we compare the three experimental set-ups considered (see Table \ref{tab:confs}) and no significant difference is found. Though some slight improvement can be seen in the low-threshold configuration (Configuration A), this happens mainly in regions that will be excluded later on after combining the results with those from other experiments. For a more detailed discussion on the impact of each different configuration on the combined analysis with JUNO, we refer the reader to Appendix \ref{sec:configurations}.

%%%%%%%%%%%%%%%%%%%%%%%%%%%%%%%%%
\subsection{NSI in JUNO}
%%%%%%%%%%%%%%%%%%%%%%%%%%%%%%%%%

In our analysis, we limit ourselves to the study of two NSI parameters simultaneously, $\varepsilon_{ee}^{dV}$ and $\varepsilon_{e\tau}^{dV}$. The motivation behind this choice is twofold: firstly, these two parameters are among the least constrained \cite{Farzan:2017xzy} and, secondly, they can be easily mapped onto the two effective parameters ($\varepsilon$ and $\varepsilon '$) used to describe NSI in solar neutrinos.

Moreover, we will assume all NSI coefficients to be real, so that
\begin{flalign}
\varepsilon ' = \sin 2\theta_{13}\cos \theta_{23} \cos\delta_{\rm CP}\varepsilon_{e\tau}^{dV} -\cos ^2 \theta_{13}\varepsilon_{ee}^{dV}\,,
\end{flalign}
and
\begin{flalign}
\varepsilon  = -\cos \theta_{13} \sin \theta_{23}\varepsilon_{e\tau}^{dV}\,.
\end{flalign}

\begin{figure}[h]
\centering
\includegraphics[width = \textwidth]{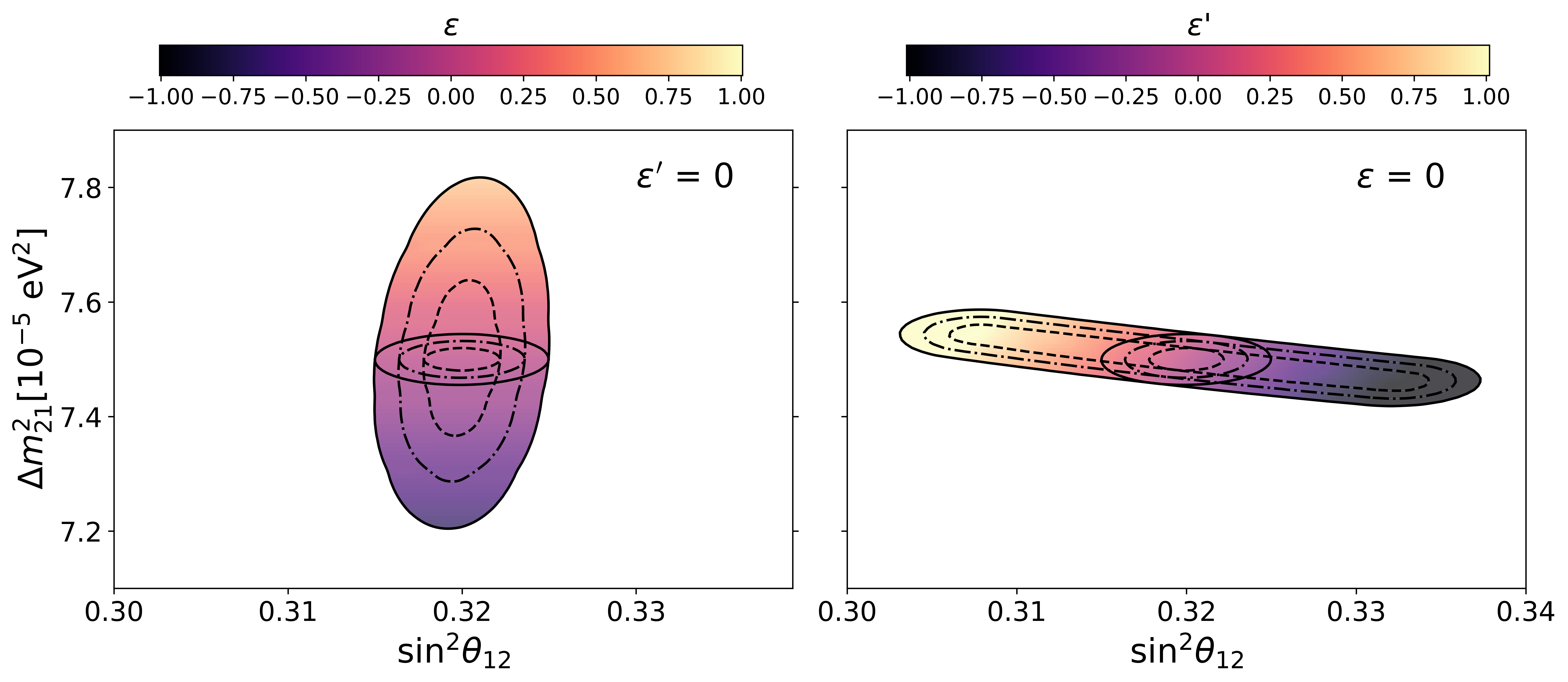}
\caption{
\label{fig:JUNOoneatatime}
Effect of the effective NSI parameters on the sensitivity of JUNO, varying one at a time. Left panel shows the 1$\sigma$, 2$\sigma$ and 3$\sigma$ C.L. in the $\sin^2 \theta_{12}- \Delta m^2_{21}$ plane when varying $\varepsilon$ between -1 and 1. Same confidence levels are drawn in the right panel for the case of $\varepsilon = 0$ and $\varepsilon '$ allowed to vary. The colour map indicates the best fit value of the effective NSI parameters. The confidence levels expected for JUNO when NSI are not included in the analysis are indicated by unfilled contours. }
\end{figure}

In a medium-baseline reactor experiment aiming to measure the oscillation parameters $\theta_{12}$ and $\Delta m^2_{21}$ with high precision, matter effects have been shown to play an important role \cite{Khan:2019doq, Li:2016txk}. In particular, matter effects produce an approximately $ 0.2\,\%$ shift in the effective mass splitting and a $1.2 \,\%$ shift in the value of effective $\sin^2 \theta_{12}$ with respect to the values which would be obtained if matter effects were not considered \cite{Khan:2019doq}. Since JUNO aims to perform a measurement of these two oscillation parameters with a precision of $\sim 0.5-0.7\,\%$, matter effects are clearly very relevant. 
Likewise, the existence of non-standard interactions, even if smaller than standard matter effects, could greatly affect the precision goals of this experiment. In Figure \ref{fig:JUNOoneatatime}, it can be seen how the sensitivity to the solar oscillation parameters is affected if the  effective NSI couplings $\varepsilon$ and $\varepsilon'$ are included in the analysis and allowed to vary between -1 and 1. For both panels, the absence of NSI was assumed as the true hypothesis, with the best fit values for the oscillation parameters taken from Table \ref{tab:oscparams}, while the test hypothesis consisted of $\varepsilon\neq0$ (left panel) and $\varepsilon'\neq0$ (right panel).

It can be seen that a non-zero $\varepsilon$ results mainly in a shift in the effective mass splitting, whereas the main impact of a non-zero $\varepsilon '$  would be a distortion in the reconstructed value of the solar mixing angle.

%

%%%%%%%%%%%%%%%%%%%%%%%%%%%%%%%%%%%%%%
\subsection{Combining Hyper-Kamiokande and JUNO}
\label{subsec:combined}
%%%%%%%%%%%%%%%%%%%%%%%%%%%%%%%%%%%%%%

The fact that the impact of NSI on solar neutrino experiments and on long and medium-baseline reactor experiments is substantially different can be exploited to further constrain such interactions. In their absence, the determination of the solar parameters $\sin^2\theta_{12}$ and $\Delta m^2_{21} $ would be heavily dominated by JUNO. However, this picture changes significantly when the possibility of NSI is accounted for.

As has been shown, the determination of oscillation parameters from medium-baseline reactor experiments alone is still quite robust, since matter effects (and other effects alike) are subdominant. Nevertheless, there would be a considerable degradation in the accuracy of the measurement itself, and so the precision goals of the experiment would not be reached. By contrast, solar neutrino experiments are very sensitive to any new physics affecting propagation and, as such, they can deliver powerful tests of neutrino interactions during propagation as long as the oscillation picture is well-established. This complementarity is what motivates the combination of both experiments as a way to set stronger bounds on non-standard interactions. 

\begin{figure}[b!]
\centering
\includegraphics[width=0.48\textwidth]{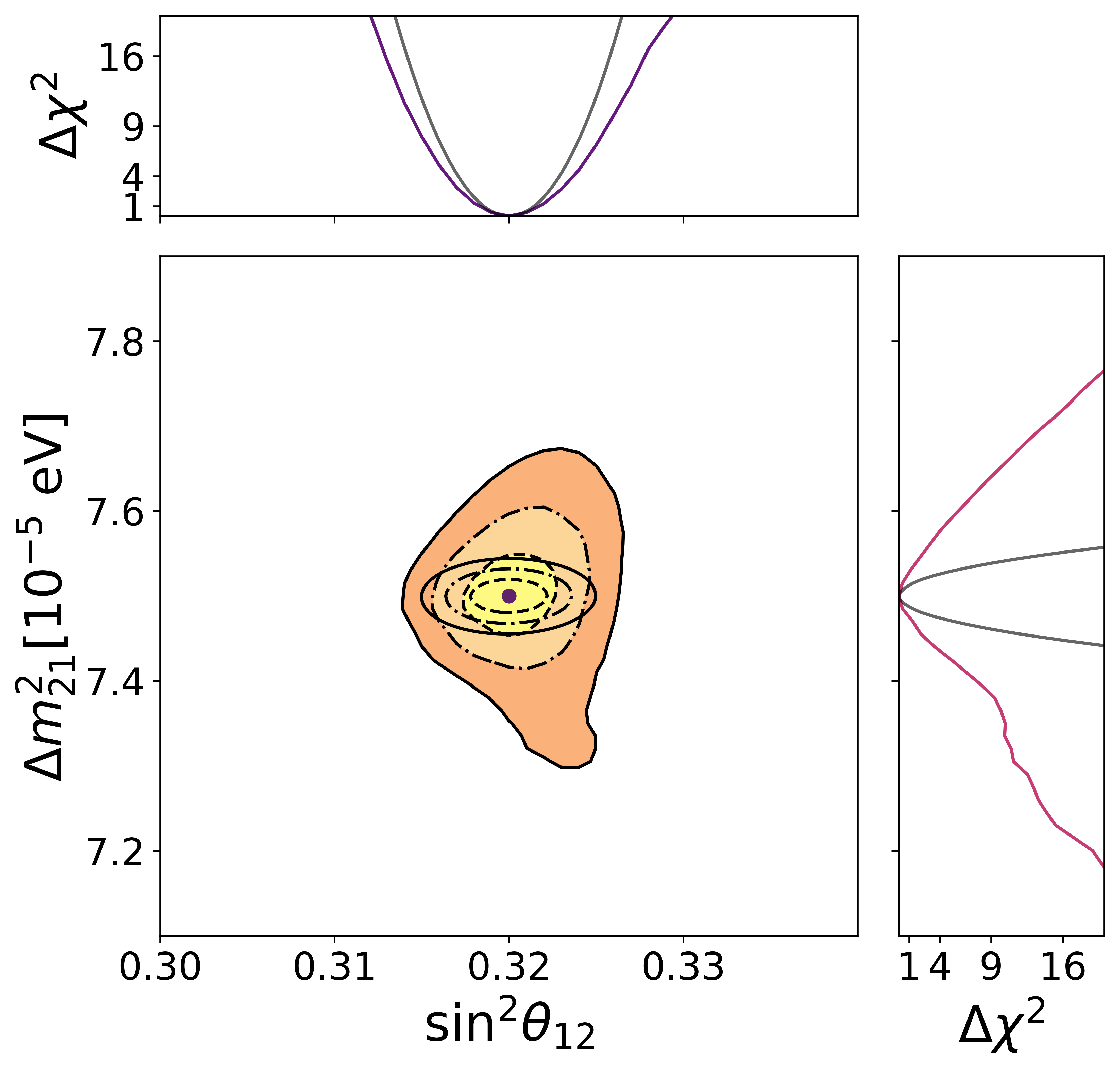}
\hspace{0.02\textwidth}
\includegraphics[width=0.48\textwidth]{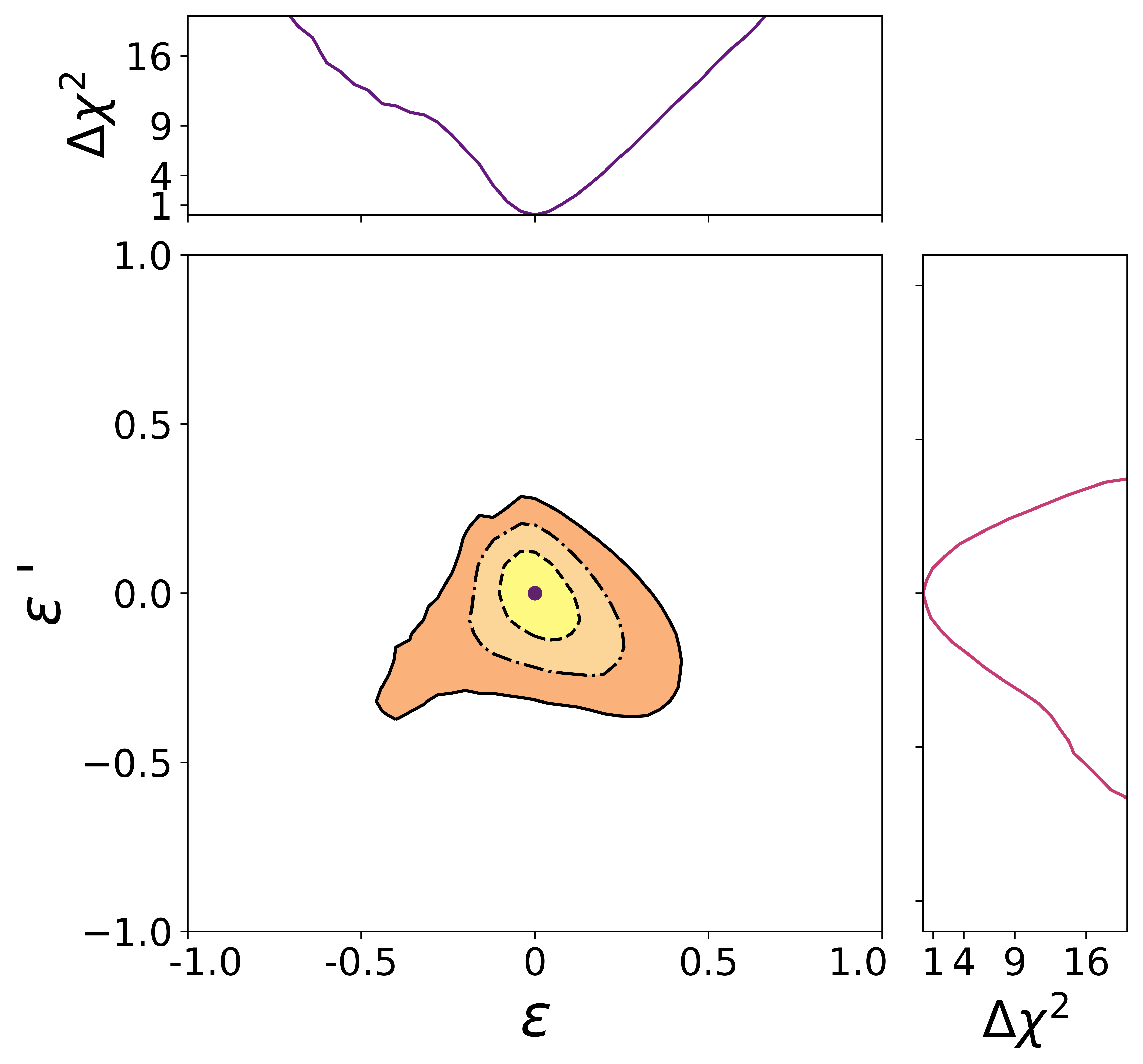}
\caption{
\label{fig:comb1stoct}
Two-dimensional projections onto the $\sin^2 \theta_{12}$ - $\Delta m^2_{21}$ plane (left panel) and the $\varepsilon$-$\varepsilon'$ plane (right panel) of the expected sensitivity from a combined analysis of Hyper-Kamiokande and JUNO. Contours correspond to 1$\sigma$ (dashed), 2$\sigma$ (dot-dashed) and 3$\sigma$ (solid) C.L. One-dimensional projections are shown for completeness. In the left panel, the corresponding confidence levels expected in the absence of NSI are again shown using unfilled contours.}
\end{figure}
\begin{figure}[b]
\centering
\includegraphics[width=0.8\textwidth]{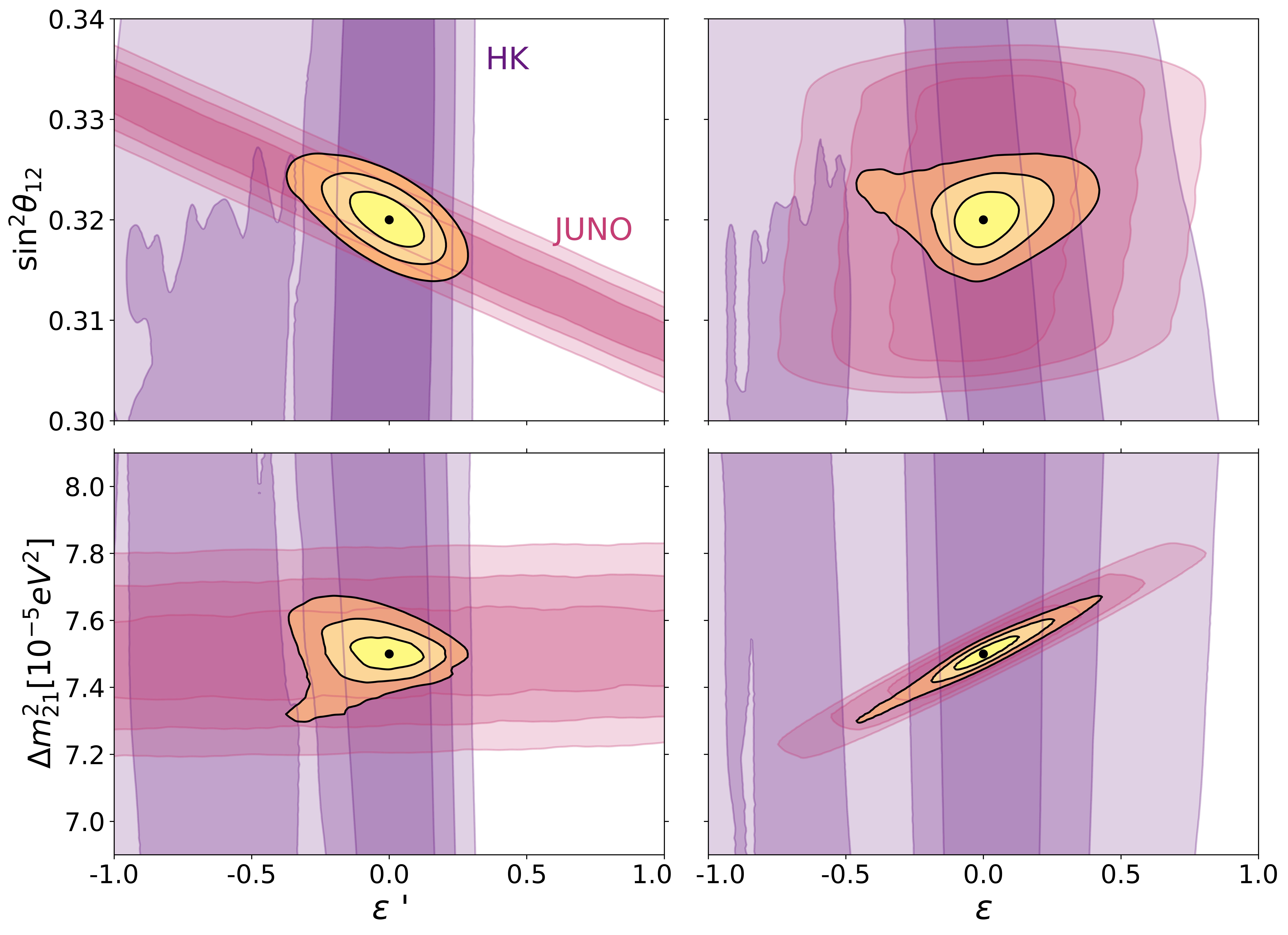}
\caption{
\label{fig:othercomb}
Two-dimensional projections onto the $\varepsilon'$ - $\sin^2 \theta_{12}$ plane (top left), $\varepsilon$ - $\sin^2 \theta_{12}$ plane (top right), $\varepsilon'$ - $\Delta m^2_{21}$ plane (bottom left) and $\varepsilon$ - $\Delta m^2_{21}$ plane (bottom right), from a combined analysis of Hyper-Kamiokande (HK) and JUNO. Contours correspond to 1$\sigma$, 2$\sigma$ and 3$\sigma$ C.L. The allowed regions from HK and JUNO individually are shown in purple and pink, respectively.}
\end{figure}

In Figure \ref{fig:comb1stoct}, we present the combined sensitivity of JUNO and Hyper-Kamiokande (using the most optimistic configuration, referred to as Configuration A in Table \ref{tab:confs}). The remaining additional projections are also shown in Figure \ref{fig:othercomb}.

In the left panel of Figure \ref{fig:comb1stoct}, it can be seen that a combination of both experiments can provide a measurement of the oscillation parameters in the solar sector reaching the subpercent precision level. At 90$\%$ C.L., the allowed regions for the solar mixing angle and mass splitting would be the following: 
\begin{equation}
0.318 < \sin^2\theta_{12} < 0.322 \quad \quad  \& \quad \quad
7.48 \times 10 ^{-5} \text{eV}^2 < \Delta m^2_{21} < 7.52 \times 10^{-5} \text{eV}^2 \,.
\end{equation}

The projected sensitivity to $\sin^2 \theta_{12}$ is very close to what JUNO alone would be expected to obtain in the absence of NSI, meaning that at least one of the oscillation parameters will be determined with this precision independently of NSI. When it comes to the solar mass splitting, however, the sensitivity will be significantly degraded if one allows for NSI, while still presenting a remarkable improvement with respect to its current level.

The right panel of Figure \ref{fig:comb1stoct} shows the projected sensitivity to NSI parameters after combining JUNO and Hyper-Kamiokande. At 90$\%$ C.L., the allowed regions for NSI parameters would be: 
\begin{equation}
    -0.153 < \varepsilon' < 0.135 \quad \quad  \& \quad \quad -0.113 < \varepsilon < 0.144 \, ,
    \label{eqn:limitsNSI}
\end{equation}

where the limits have been obtained by varying one parameter at a time. It should be noted that large values of $\varepsilon'$ are excluded - this is due to the fact that we are assuming the same mass ordering for the true values and the ones being tested. In this case, the LMA-D solution, which is only possible for large $\varepsilon'$ and different orderings for each set of values, does not arise. This approach is justified if the mass ordering is determined independently of NSI; in the next section, we will relax this constraint and study the case in which the mass orderings are allowed to be different.

The individual constraints from Hyper-Kamiokande (HK) and JUNO are shown in order to illustrate that the sensitivity to NSI parameters arises from the combination of both experiments. As mentioned previously, JUNO's sensitivity to the oscillation parameters is not greatly affected when NSI are included in the analysis; conversely, small non-standard interactions would induce large deviations and a significant loss of accuracy in the measurement of neutrino oscillation parameters by Hyper-Kamiokande. Figure \ref{fig:othercomb} captures these two complementary features.

%%%%%%%%%%%%%%%%%%%%%%%%%%%%%%%%%%%%%%%%%%%%%%%%%%%%
\subsection{LMA-D in the case of inverted ordering}
\label{subsec:dark}
%%%%%%%%%%%%%%%%%%%%%%%%%%%%%%%%%%%%%%%%%%%%%%%%%%%%

\begin{figure}[h!]
\centering
\includegraphics[width=0.48\textwidth]{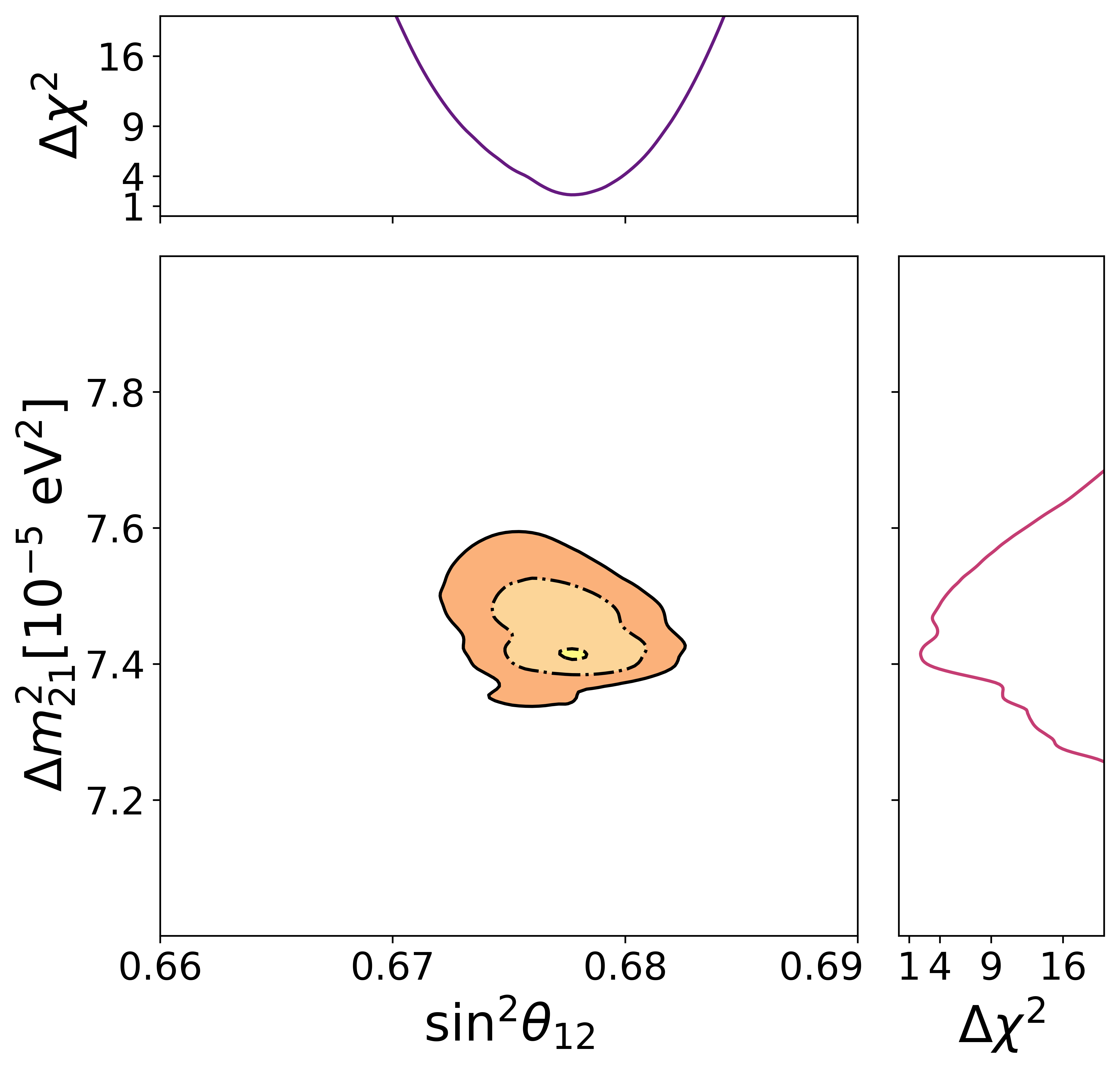}
\hspace{0.02\textwidth}
\includegraphics[width=0.48\textwidth]{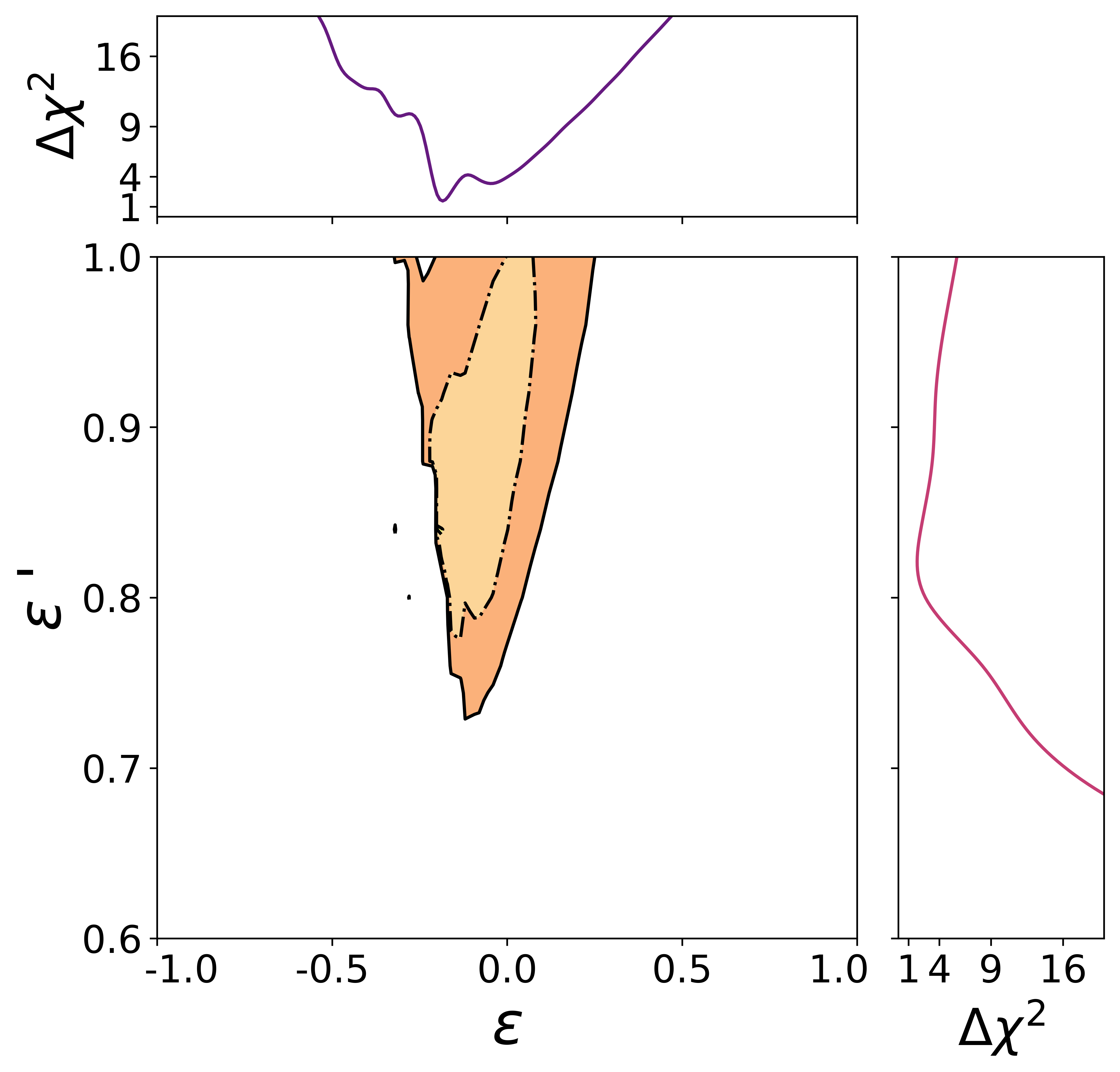}
\caption{Two-dimensional projections onto the $\sin^2 \theta_{12}$ - $\Delta m^2_{21}$ plane (left panel) and the $\varepsilon$ - $\varepsilon'$ plane (right panel)  of the expected sensitivity from a combined analysis of Hyper-Kamiokande and JUNO for the LMA-D solution. Contours correspond to 1$\sigma$ (dashed), 2$\sigma$ (dot-dashed) and 3$\sigma$ (solid) C.L. with respect to the best fit in the first octant from Table \ref{tab:oscparams} and assuming normal ordering. One-dimensional projections are shown for completeness.}
\label{fig:comb_2ndoct}
\end{figure}

Up until this point, we have assumed that the mass ordering would be determined in an NSI-independent way, or at least that a strong preference for a particular ordering would be achieved when the analysis was extended to include NSI. If this limiting assumption is lifted, a second solution arises - known as the LMA-D solution, it is a consequence of the generalised mass degeneracy as explained in Section \ref{sec:teo}.
The allowed regions for this solution in the $\sin^2\theta_{12}-\Delta m^2_{21}$ and $\varepsilon - \varepsilon'$ planes are shown in the left and right panels of Figure~\ref{fig:comb_2ndoct}, respectively.

In the analysis of solar neutrinos, there is a negligible dependence on $\Delta m^2_{31}$ and its sign. Therefore, in Hyper-Kamiokande, the correct methodology to study the degenerate \mbox{LMA-D} solution with $\sin^2\theta_{12} > 0.5$  is essentially identical to the analysis described in the previous section.
In the case of JUNO, however, there is a strong dependence on $\Delta m^2_{31}$, and by fixing its value to its best fit point under normal mass ordering we were systematically excluding this possibility. Therefore, the appropriate procedure for exploring the degenerate solution in JUNO is to reconstruct the ``mock data" (which assumes no NSI and a best fit for $\sin ^2\theta_{12} < 0.5$, i.e. in the first octant) while allowing $\Delta m^2_{31}$ to take both positive and negative values (thus accounting for both hierarchies).

In this case, a second solution arises in the second octant, that is, for $\sin^2 \theta_{12} > 0.5$. The best fit for this degenerate solution is slightly disfavoured with respect to the one in the first octant, which can be seen from the fact that $\Delta \chi^2$ is larger than zero for this solution. This is because the number density of $d$-quarks differs for each medium (the Earth's crust, core and within the Sun), and different values of $\varepsilon'$ can therefore account for the generalised mass ordering degeneracy in each medium (see Eq.~(\ref{eqn:epsdark2})). Nevertheless, in a combined analysis of JUNO and Hyper-Kamiokande, the LMA-D solution would not be discounted. This solution also arises for a slightly smaller $\Delta m^2_{21}$, as seen in \mbox{Figure \ref{fig:comb_2ndoct}}. 

After marginalising over the solar neutrino oscillation parameters, one can obtain the sensitivity of the combined analysis to the effective NSI parameters, $\varepsilon$ and $\varepsilon'$, as shown in the right panel of Figure \ref{fig:comb_2ndoct}. It is clear from this figure that the LMA-D region requires very large values of $\varepsilon'$, as well as a non-zero $\varepsilon$, which is in agreement with previous works~\cite{Miranda:2004nb,Escrihuela:2009up}. 
In spite of the magnitude of these values, neutrino oscillation data alone would not be able to exclude them. However, scattering data and results from coherent elastic neutrino-nucleus scattering  experiments~\cite{COHERENT:2017ipa} in particular have been shown to be a powerful complementary probe for this scenario~\cite{Coloma:2017ncl,Esteban:2018ppq,Coloma:2019mbs}. Indeed, the combination of solar neutrino data with results from the COHERENT experiment leads to a rejection of the LMA-D solution above the 3$\sigma$ level in models with NSI involving only a single quark flavour~\cite{Coloma:2017ncl, Akimov:2021dab}. Nevertheless, this constraint is  relaxed when allowing for NSI with both $u$ and $d$-type quarks simultaneously, so it is not possible to fully exclude the LMA-D solution in these more generic scenarios~\cite{Esteban:2018ppq, Coloma:2019mbs}.

\begin{figure}[t!]
\centering
\includegraphics[width=0.8\textwidth]{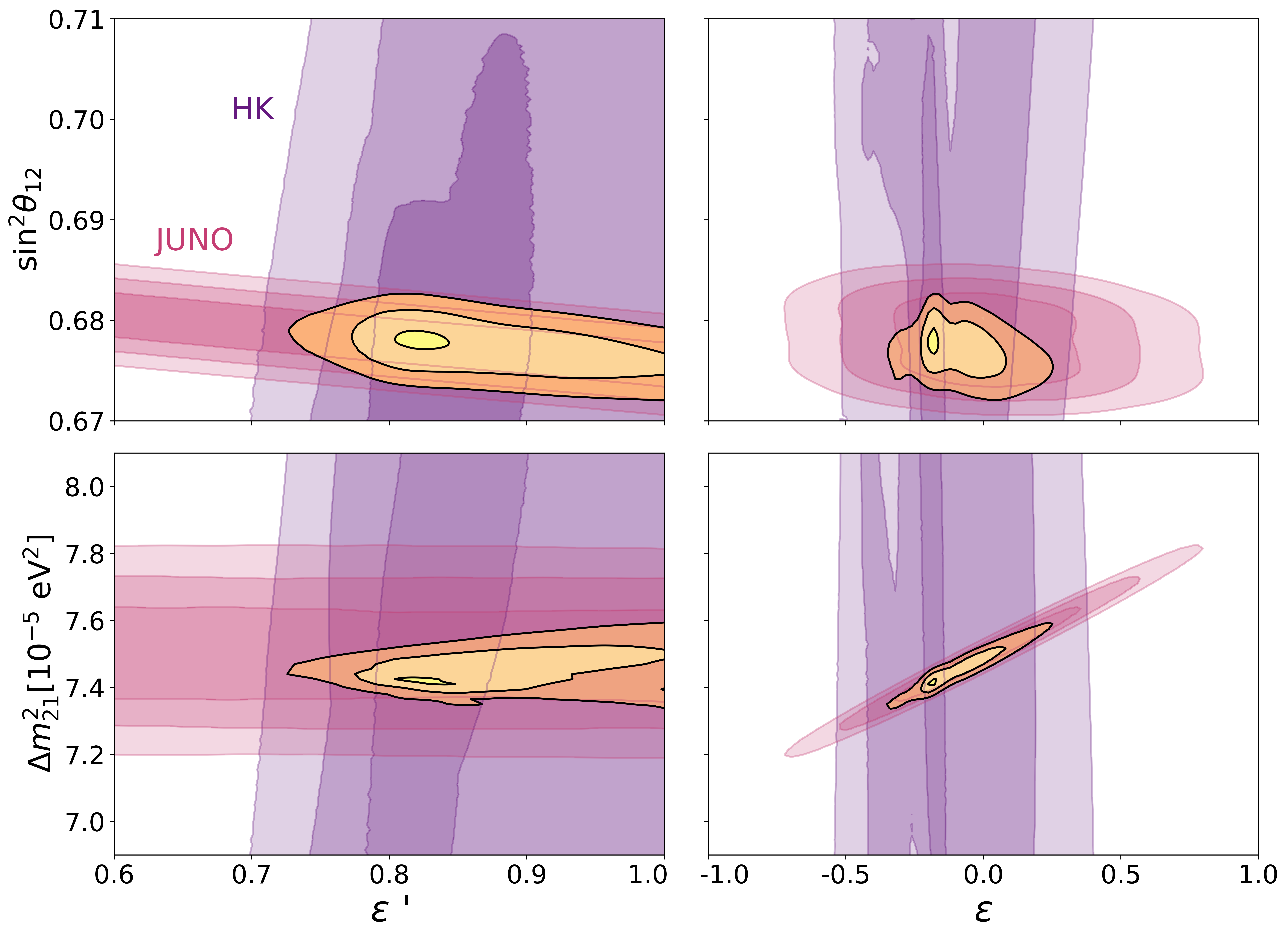}
\caption{Two-dimensional projections onto the $\varepsilon'$ - $\sin^2 \theta_{12}$ plane (top left), $\varepsilon$ - $\sin^2 \theta_{12}$ plane (top right), $\varepsilon'$ - $\Delta m^2_{21}$ plane (bottom left) and $\varepsilon$ - $\Delta m^2_{21}$ plane (bottom right), from a combined analysis of Hyper-Kamiokande (HK) and JUNO, for the LMA-D solution. Contours correspond to 1$\sigma$, 2$\sigma$ and 3$\sigma$ C.L. The allowed regions from HK and JUNO individually are shown in purple and pink, respectively.  }
\label{fig:other2ndoct}
\end{figure}

Finally, and for completeness, we show in Figure \ref{fig:other2ndoct} the projections onto the remaining planes for the LMA-D solution. Once again, the sensitivity obtained for JUNO and Hyper-Kamiokande individually is shown together with the resulting sensitivity from a combined analysis. 

%%%%%%%%%%%%%%%%%%%%%%%%%%%%%%%%%%%%%%%%%%%%%%
%%%%%%%%%%%%%%%%%%%%%%%%%%%%%%%%%%%%%%%%%%%%%%
\section{Conclusions}
\label{sec:conclusion}
%%%%%%%%%%%%%%%%%%%%%%%%%%%%%%%%%%%%%%%%%%%%%%
%%%%%%%%%%%%%%%%%%%%%%%%%%%%%%%%%%%%%%%%%%%%%%

The complementarity between solar and reactor experiments is known to be particularly powerful when it comes to constraining neutrino non-standard interactions. We have addressed the expected improvements which may be achieved from a combined analysis of the future neutrino experiments JUNO and Hyper-Kamiokande, focusing on NSI with $d$-type quarks. 

Including non-standard interactions in the analysis of JUNO data would result in a significantly degraded sensitivity to the solar oscillation parameters $\sin^2\theta_{12}$ and $\Delta m^2_{21}$. Nonetheless, a combined analysis with Hyper-Kamiokande would allow a subpercent precision measurement of $\sin^2\theta_{12}$ and $\Delta m^2_{21}$ at 90\% confidence level.

We have shown that the results obtained are not strongly dependent on the exact experimental setup used for Hyper-Kamiokande, such as the fiducial volume or the energy threshold. This is a consequence of the combined analysis relying on a more precise determination of the oscillation parameters by JUNO. Nevertheless, a better determination of the day-night asymmetry and an observation of the upturn in the solar neutrino spectrum remains key for verifying our understanding of the solar neutrino picture.

If independent probes were able to exclude large values of $\varepsilon'$, the limits derived in this work and shown in Eq.~\eqref{eqn:limitsNSI} would also be the only allowed ranges from the combination of JUNO and Hyper-Kamiokande data for the effective NSI parameters. These constraints, which are comparable to similar sensitivity studies for future neutrino experiments \cite{Liao:2017awz, Bakhti:2020fde}, will improve the current bounds from combined analyses of solar and KamLAND data~\cite{Miranda:2004nb,Escrihuela:2009up}, as expected. In the same spirit as  
global neutrino oscillation fits including NSI~\cite{Esteban:2018ppq,Coloma:2019mbs}, a combined fit of Hyper-Kamiokande, JUNO and future results from coherent elastic neutrino-electron scattering will exploit the complementary sensitivity of these three types of experiments to NSI, providing crucial information on the nature of neutrino interactions with matter.

Finally, we explored the possibility of constraining the LMA-D solution from the difference in matter effects between Hyper-Kamiokande and JUNO. We have shown that only a very small region of the parameter space would be allowed at 1$\sigma$. However, as expected, it would not be possible to completely exclude this solution from the combination of these two experiments alone.

\section*{Acknowledgements}
P.\,M.\,M.\, is grateful for the hospitality of the Particle and Astroparticle Physics Division of the \mbox{Max-Planck-Institut} f{\"u}r Kernphysik (Heidelberg) during the last stages of this project.
This work has been supported by the Spanish grants PID2020-113775GB-I00 (AEI/10.13039/501100011033) and PROMETEO/2018/165 (Generalitat Valenciana).
P.\,M.\,M.\, is also supported by grant FPU18/04571.

\appendix

\section{On the possibility of different Hyper-Kamiokande configurations}
\label{sec:configurations}

\begin{figure}[b!]
\centering
\includegraphics[width=0.8\textwidth]{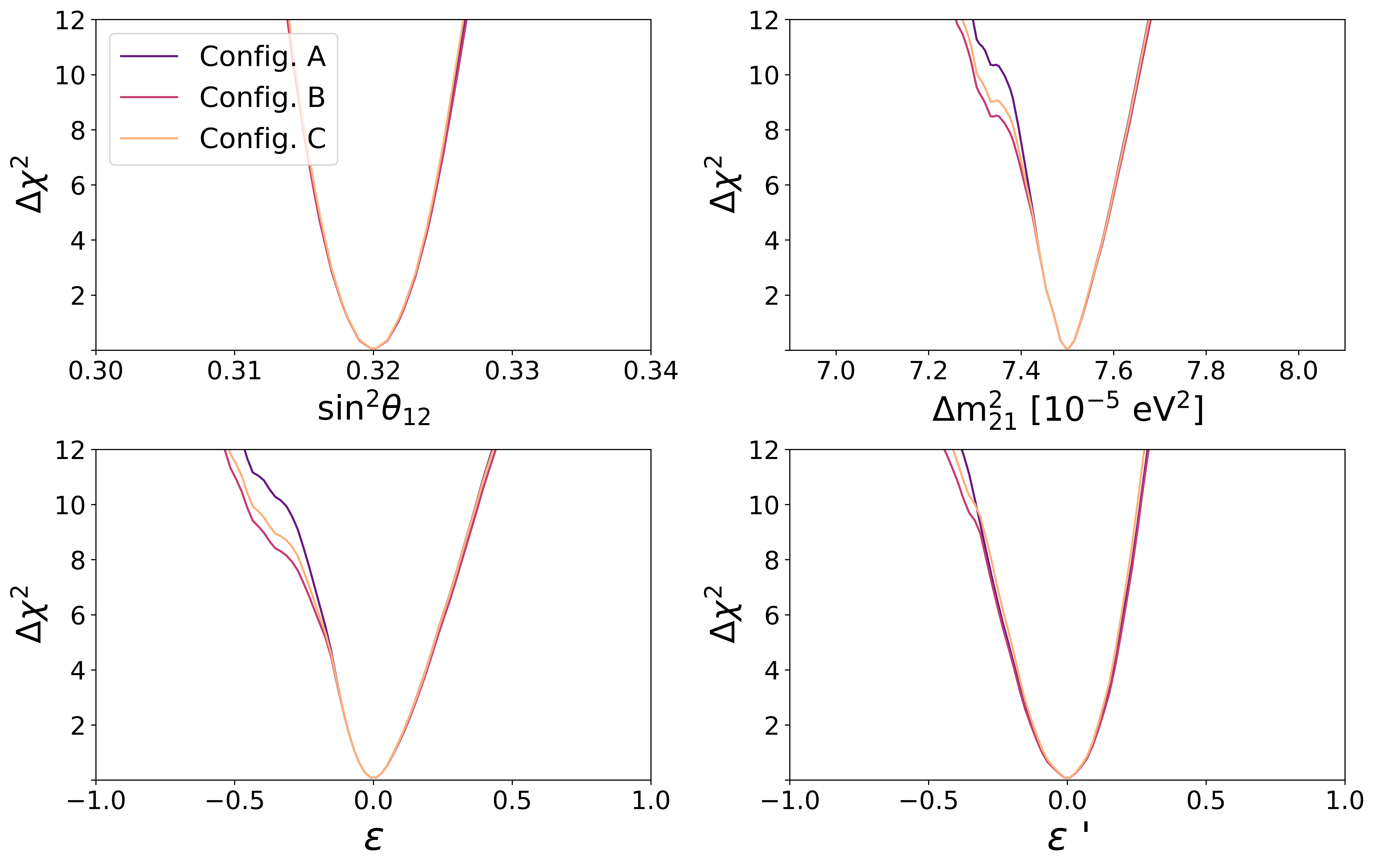}
\caption{
Sensitivity to neutrino oscillation and NSI parameters for the three different potential configurations of the Hyper-Kamiokande detector considered in this work.}
\label{fig:chi2_profiles}
\end{figure}

In Section~\ref{sec:NSI-HK} we showed the impact of three possible experimental configurations of Hyper-Kamiokande on its determination of solar oscillation parameters in the presence of NSI (see Figure~\ref{fig:HK-confs}).
It is also interesting to examine whether our final results depend substantially on the exact configuration of the Hyper-Kamiokande detector. 
The sensitivity profiles of each of the four parameters under consideration - $sin^2\theta_{12}$, $\Delta m^2_{21}$, $\varepsilon$ and $\varepsilon'$ - are presented in Figure \ref{fig:chi2_profiles}.

It can be clearly seen that the choice of one experimental set-up over another would not have a significant impact on the determination of the oscillation parameters $\sin^2 \theta_{12}$ and $\Delta m^2_{21}$. This is due to the fact that, in the presence of NSI, there are no notable differences in the sensitivity of Hyper-Kamiokande, at least in the region of parameter space where the measurement of JUNO will sit.
The only noticeable divergent behaviour arises at the $\sim 2-3\,\sigma$ level for the low-$\Delta m^2_{21}$ and low-$\varepsilon$ sides of those profiles, respectively, which corresponds to the lobules which appear in the two-dimensional regions presented in Section \ref{sec:results}. Thus, the biggest differences between the three configurations listed in Table \ref{tab:confs} appear for values that will be excluded by JUNO, as shown in Figure \ref{fig:HK-confs}.

\bibliography{bibliography}

\end{document}